\def\zem{$z_{\rm em}$}
\def\zabs{$z_{\rm abs}$}

\def\cm2{cm$^{-2}$}
\def\loghi{\mbox{$\log N_{\rm HI}$}}
\def\nhi{\mbox{$N_{\rm HI}$}}
\def\hi{H~{\sc i}}

\def\cii{C~{\sc ii}  $\lambda$ }
\def\civ{C~{\sc iv}  $\lambda$ }
\def\feii{Fe~{\sc ii} $\lambda$ }
\def\feiii{Fe~{\sc iii} $\lambda$ }
\def\mg1{Mg~{\sc i}}
\def\mg2{Mg~{\sc ii}}
\def\sii{S~{\sc ii} $\lambda$ }
\def\siii{Si~{\sc ii} $\lambda$ }
\def\siiv{Si~{\sc iv} $\lambda$ }
\def\alii{Al~{\sc ii} $\lambda$ }
\def\aliii{Al~{\sc iii} $\lambda$ }
\def\znii{Zn~{\sc ii} $\lambda$ }
\def\oi{O~{\sc i} $\lambda$ }
\def\o6{O~{\sc vi}}
\def\n1{N~{\sc i}}
\def\h1{H~{\sc i}}

\def\approxlt{\mathrel{\spose{\lower 3pt\hbox{$\sim$}}
        \raise 2.0pt\hbox{$<$}}}
\def\approxgt{\mathrel{\spose{\lower 3pt\hbox{$\sim$}}
        \raise 2.0pt\hbox{$>$}}}

\documentstyle[psfig,rotating]{mn}

\input{psfig.sty}

\newif\ifAMStwofonts

\title[Global Metallicity Evolution]{A Homogeneous Sample of Sub-DLAs IV: \hspace{4cm} Global Metallicity Evolution\thanks{Based on observations collected during programme ESO 71.A-0114, ESO 73.A-0653, ESO 73.A-0071 and ESO 74.A-0306 at the European Southern Observatory with UVES on the 8.2 m KUEYEN telescope operated at the Paranal Observatory, Chile.}}

\author[C\'eline P\'eroux et al.]
       {C\'eline P\'eroux$^{1}$\thanks{e-mail: cperoux@eso.org},
       Miroslava Dessauges-Zavadsky$^{2}$, Sandro D'Odorico$^{1}$,
\newauthor
Tae-Sun Kim$^{3}$ \& Richard G. McMahon$^{4}$\\
$1$ European Southern Observatory, Karl-Schwarzchild-Str. 2, 85748 Garching-bei-M\"unchen, Germany.\\
$2$ Observatoire de Gen\`eve, 1290 Sauverny, Switzerland.\\
$3$ Astrophysikalisches Institut Potsdam, An der Sternwarte 16, 14482 Potsdam, Germany.\\
$4$ Institute of Astronomy, Madingley Road, Cambridge CB3 0HA, UK.\\
}

\date{}

\pagerange{\pageref{firstpage}--\pageref{lastpage}}
\begin{document}

\maketitle

\label{firstpage}

\begin{abstract}
An accurate method to measure the abundance of high-redshift galaxies
consists in the observation of absorbers along the line of sight
toward a background quasar. Here, we present abundance measurements of
13 $z\geq3$ sub-Damped Lyman-$\alpha$ Systems (quasar absorbers with
\hi\ column density $19$ $<$ log N(HI) $<$ 20.3 cm$^{-2}$) based on
the high resolution observations with VLT UVES spectrograph. These
observations more than double the metallicity information for sub-DLAs
previously available at $z>3$. This new data, combined with other
sub-DLA measurements from the literature, confirm the stronger
metallicity redshift evolution than for the classical Damped
Lyman-$\alpha$ absorbers. Besides, these observations are used to
compute for the first time the fraction of gas ionised from
photo-ionisation modelling in a {\it sample} of sub-DLAs.
Based on these results, we calculate that sub-DLAs contribute no more
than 6\% of the expected amount of metals at z$\sim$ 2.5. We therefore
conclude that even if sub-DLAs are found to be more metal-rich than
classical DLAs, they are insufficient to close the so-called ``missing
metals problem''.
\end{abstract}


\section{Introduction}

Damped Lyman-$\alpha$ systems (hereafter DLAs) seen in absorption in
the spectra of background quasars are selected over all redshifts
independent of their intrinsic luminosity. They have hydrogen column
densities, log $N({\rm H I}) \geq 20.3$. DLAs are also contributors to
the neutral gas, $\Omega_{\rm HI}$, in the Universe at high redshifts
and it is from this gas reservoir that the stars visible today formed
(Wolfe et al. 1995). Furthermore, DLAs offer a direct and accurate
probe of elemental abundances over $ > 90 \%$ of the age of the
Universe.

Recently, much attention has been drawn to the sub-Damped
Lyman-$\alpha$ Systems. These systems have \hi\ column density $19$
$<$ log N(HI) $<$ 20.3 cm$^{-2}$ and were first coined by P\'eroux et
al. (2003a). In a series of paper, our group has analysed a unique
homogeneous sample of sub-DLAs all observed at the same resolution
with the same instrument on the Very Large Telescope (VLT). The first
two papers are based on the ESO/UVES archives (Paper I:
Dessauges-Zavadsky et al. 2003; Paper II: P\'eroux et al. 2003b) while
more recent studies are from our own observational programmes (Paper
III: P\'eroux et al. 2005; Paper IV: this paper) leading to a new
sample of high-redshift sub-DLAs observed under the same
conditions. These works as well as other recent studies are suggesting
that high metallicities can be found more easily in sub-DLAs than in
classic DLAs, especially a low-redshift (e.g., Pettini et al. 2000;
Jenkins et al. 2005; P\'eroux et al. 2006a; Prochaska et al. 2006;
P\'eroux et al. 2006b). In the present paper, we study a new sample of
high-redshift sub-DLAs to better constrain their metallicities at
$z>3$. In addition, our data are used in combination with others from
the literature to compute the fraction of ionised gas in a sample of
sub-DLAs and reliably compute the contribution of sub-DLAs to the
global metallicity.

The paper is structure as follows. 
In the second section, we present the methodology and results of the
determination of the chemical content of these high-redshift
sub-DLAs. In the third section, the total abundances including results
from detailed photo-ionisation models are given and the redshift
evolution of the metallicity is presented. Finally, the last section
describes the contribution of sub-DLAs to the global metallicity in
the context of the so-called missing metals problem.

%%%%%%%%%%%%%%%%%%%%%%%%%%%%%%%%%%%%%%%%%%%%%%%%%%%%%%%%%%%%%%%%%%%%%%%%
%%%%%%%%%%%%%%%%%%%%%%%%%%%%%%%%%%%%%%%%%%%%%%%%%%%%%%%%%%%%%%%%%%%%%%%%
\section{A Sample of 13 $z\geq3$ Sub-DLAs}

Table~\ref{t:Q} summarises the absorption redshift and \nhi\ column
densities for each of the 13 $z\geq3$ sub-DLAs which constitute the
sample under study. The data acquisition and reduction are described
in P\'eroux et al. (2005). In the case of PSS J0133$+$0400, our own
data have been supplemented by a 3000 sec exposure with setting 860
(ESO 74.A-0306, P.I.: Valentina D'Odorico) and a 5200 sec exposure
with setting 540 (ESO 73.A-0071, P.I.: C\'edric Ledoux).

In this section, details on each of the studied sub-DLAs are
provided. The metals associated with the Lyman lines were searched
over the spectral coverage available. More than the 40 transitions
most frequently detected in high column density quasar absorbers have
been systematically looked for. The metal column densities are then
determined by fitting Voigt profiles to the absorption lines. The fits
were performed using the $\chi^2$ minimisation routine {\tt fitlyman}
in {\tt MIDAS} (Fontana \& Ballester 1995). 

\begin{table}
\begin{center}
\caption{The sample of 13 high-redshift sub-DLAs for which abundance studies 
have been undertaken.\label{t:Q}}
\begin{tabular}{lccc}
\hline
Quasar &\zem &\zabs &\loghi \\
\hline
PSS J0118$+$0320        &4.230  &4.128  &20.02$\pm$0.15\\
PSS J0121$+$0347	&4.127  &2.976  &19.53$\pm$0.10\\
SDSS J0124$+$0044	&3.840  &2.988  &19.18$\pm$0.10\\
...                     &...    &3.078  &20.21$\pm$0.10\\
PSS J0133$+$0400        &4.154  &3.139	&19.01$\pm$0.10\\
...                     &...	&3.995	&19.94$\pm$0.15\\
...                     &...	&3.999  &19.16$\pm$0.15\\
...                     &...    &4.021	&19.09$\pm$0.15\\
BRI J0137$-$4224	&3.970  &3.101	&19.81$\pm$0.10\\
...                     &...    &3.665	&19.11$\pm$0.10\\
BR J2215$-$1611		&3.990  &3.656  &19.01$\pm$0.15\\
...                     &...    &3.662  &20.05$\pm$0.15\\
BR J2216$-$6714		&4.469  &3.368  &19.80$\pm$0.10\\
\hline			
\end{tabular}
\end{center}
\vspace{0.2cm}
\begin{minipage}{80mm}

\end{minipage}
\end{table}

\begin{enumerate}

\vspace{0.5cm}
\item{{\bf PSS J0118$+$0320 (\zem=4.230, \zabs=4.128, \loghi=20.02$\pm$0.15):}
 
This system is at the high end of the sub-DLA definition with
\loghi=20.02$\pm$0.15. Many metal lines are detected at \zabs=4.128, some of
which are clearly saturated. The fit is done using simultaneously \feii
1608, \feii 1144, \siii 1526, \siii 1304, \sii 1259 and \sii 1253 to derive
the redshifts and $b$ parameters of the 5 components. The fit is shown
in Figure~\ref{f:Q0118p0320z4p128_low} and the matching parameters are
presented in Table~\ref{t:Q0118p0320z4p128_low}. In this and the
following tables velocities and $b$ are in km~s$^{-1}$ and N's are in
cm$^{-2}$. Lower limits for $\rm C~{\sc ii}$ and $\rm O~{\sc i}$ are derived from the saturated
lines. The \znii 2026 line is covered by our spectrum but is situated
in a region too noisy to allow to derive any meaningful upper limit.

Both $\rm Si~{\sc iv}$ and $\rm C~{\sc iv}$ high-ionisation doublets are detected. The redshifts
and $b$ of the 9 components are derived from a simultaneous
fit of \civ 1548, \civ 1550 and \siiv 1393 (\siiv 1402 is affected by
telluric lines). The resulting fit is overplotted on \siiv 1402 for
consistency check. The fit is shown in
Figure~\ref{f:Q0118p0320z4p128_high} and the matching parameters are
presented in Table~\ref{t:Q0118p0320z4p128_high}.}

\begin{table*}
\begin{center}
\caption{Parameters fit to the low-ionisation transitions of the 
\zabs=4.128 \loghi=20.02$\pm$0.15 sub-DLA towards PSS
J0118$+$0320. \label{t:Q0118p0320z4p128_low}}
\begin{tabular}{r r r r r r r r r r r}
\hline
z & $b$ &log N($\rm Fe~{\sc ii}$) &log N($\rm Si~{\sc ii}$) &log N($\rm C~{\sc ii}$) &log N($\rm O~{\sc i}$) &log N($\rm S~{\sc ii}$) \\
\hline
4.127836&  4.61$\pm$0.14&13.21$\pm$0.11&13.60$\pm$0.02&$>$15.35&$>$13.95 &13.41$\pm$0.11\\    
4.128319& 30.06$\pm$0.21&13.57$\pm$0.12&14.14$\pm$0.02&$>$14.58&$>$15.15 &13.94$\pm$0.11\\   
4.128678&  3.11$\pm$0.14&13.71$\pm$0.11&14.69$\pm$0.04&$>$17.28&$>$16.37 &13.75$\pm$0.11\\   
4.129264&  3.30$\pm$0.18&13.43$\pm$0.12&13.42$\pm$0.02&$>$15.64&$>$14.02 &13.02$\pm$0.13\\   
4.129627&  6.87$\pm$0.25&13.12$\pm$0.11&12.73$\pm$0.01&$>$12.60&$>$12.00 &11.89$\pm$0.12\\   
\hline 				       			 	  
\end{tabular}			       			 	  
\end{center}			       			 	  
\end{table*}			       			 	  

\begin{figure}
\begin{center}
\includegraphics[height=10cm, width=8cm, angle=0]{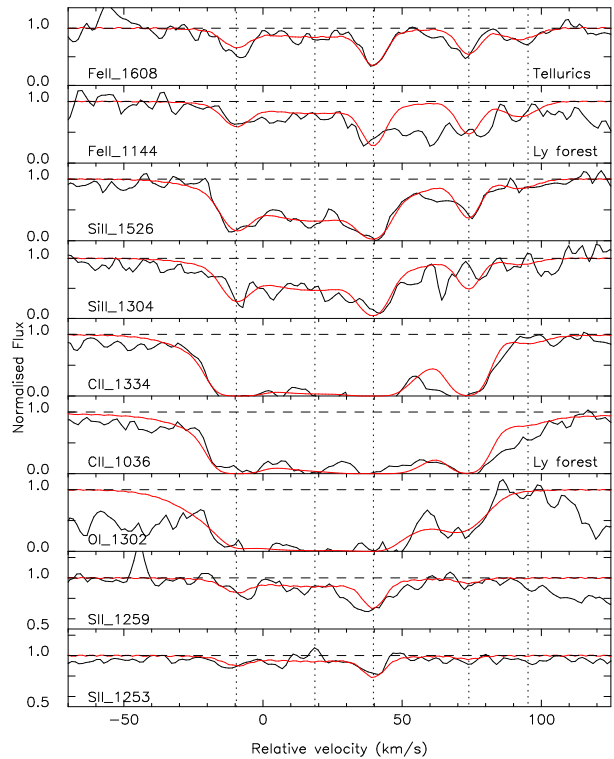}
\caption{Fit to the low-ionisation transitions of the 
\zabs=4.128 \loghi=20.02$\pm$0.15 sub-DLA towards PSS
J0118$+$0320 (see Table~\ref{t:Q0118p0320z4p128_low}). In this and 
the following figures, the zero velocity
corresponds to the absorption redshift listed in Table~\ref{t:Q}, the
vertical dashed lines correspond to the fitted components and the
mention 'Ly forest' means that the metal line falls in the
Lyman-$\alpha$ forest of the quasar spectrum while 'Tellurics'
indicates a region contaminated by telluric
lines. \label{f:Q0118p0320z4p128_low}}
\end{center}
\end{figure}
    
\begin{table}
\begin{center}
\caption{Parameters fit to the high-ionisation transitions of the 
\zabs=4.128 \loghi=20.02$\pm$0.15 sub-DLA towards PSS
J0118$+$0320.\label{t:Q0118p0320z4p128_high} }
\begin{tabular}{ r r r r }
\hline
z      & $b$    &log N($\rm C~{\sc iv}$) &log N($\rm Si~{\sc iv}$)\\ 
\hline
4.126680&   29.70$\pm$0.26&13.49$\pm$0.12&   13.17$\pm$0.12\\    
4.127345&    9.90$\pm$0.13&13.19$\pm$0.12&   12.47$\pm$0.12\\   
4.128024&   22.20$\pm$0.31&13.30$\pm$0.11&   12.71$\pm$0.22\\   
4.128656&   11.10$\pm$0.12&13.30$\pm$0.33&   12.56$\pm$0.12\\   
4.129083&   21.60$\pm$0.23&13.59$\pm$0.11&   13.04$\pm$0.11\\   
4.129787&    3.20$\pm$0.02&13.04$\pm$0.12&   12.30$\pm$0.12\\   
4.130161&    8.70$\pm$0.04&13.50$\pm$0.11&   12.97$\pm$0.12\\   
4.130446&    2.60$\pm$0.01&13.08$\pm$0.12&   12.65$\pm$0.12\\   
4.130711&    9.00$\pm$0.08&13.31$\pm$0.11&   12.84$\pm$0.12\\   
\hline 				       			 	  
\end{tabular}			       			 	  
\end{center}			       			 	  
\end{table}

\begin{figure}
\begin{center}
\includegraphics[height=6cm, width=8cm, angle=0]{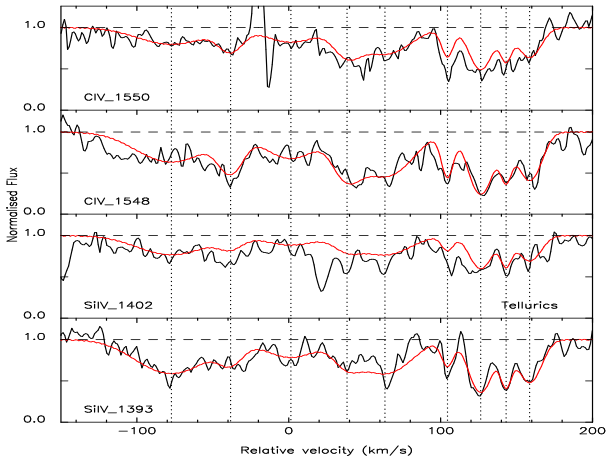}
\caption{Fit to the high-ionisation transitions of the 
\zabs=4.128 \loghi = 20.02$\pm$0.15 sub-DLA towards PSS
J0118$+$0320 (see Table~\ref{t:Q0118p0320z4p128_high}).\label{f:Q0118p0320z4p128_high}}
\end{center}
\end{figure}

%%%%%%%%%%%%%%%%%%%%%%%%%%%%%%%%%%%%%%%%%%%%%%%%%%%%%%%%%%%%%%%%%%%%%%%%%%%%%%%%%%%%%%%%%%%%%%%%%

\vspace{0.5cm}
\item{{\bf PSS J0121$+$0347 (\zem=4.127, \zabs=2.976, \loghi=19.53$\pm$0.10):}

The low-ionisation transitions in this system are well fitted with one
component, the redshift and $b$ of which are fixed by a simultaneous
fit of \siii 1260, \oi 1302, \cii 1334, \feii 1608 and \alii 1670. The
first four lines are situated in the Lyman-$\alpha$ forest. This
explains the likely blending which affects them. Therefore,
measurements from  $\rm Si~{\sc iii}$, $\rm O~{\sc i}$ and $\rm C~{\sc ii}$ 
are considered as upper limits. The
fit is shown in Figure~\ref{f:Q0121p0347z2p976_low} and the matching
parameters are presented in Table~\ref{t:Q0121p0347z2p976_low}. \aliii
1854 and \aliii 1862 are covered but not detected. An upper limit is
derived for the Al abundance: log N($\rm Al~{\sc iii}$)$<$11.77 at 4$\sigma$.

Interestingly, in this system the high-ionisation transitions is
partially fitted with the same velocity component than for the
low-ionisation transitions. The $\rm C~{\sc iv}$ and $\rm Si~{\sc iv}$ doublets are fitted
simultaneously. $\rm C~{\sc iv}$ is heavily blended and therefore leads to an upper
limit, while $\rm Si~{\sc iv}$ is nicely fitted from the \siiv 1393 line. The fit is
shown in Figure~\ref{f:Q0121p0347z2p976_high} and the matching
parameters are presented in Table~\ref{t:Q0121p0347z2p976_high}.

}

\begin{table*}
\begin{center}
\caption{Parameters fit to the low-ionisation transitions of the 
\zabs=2.976 \loghi=19.53$\pm$0.10 sub-DLA towards PSS J0121$+$0347. 
\label{t:Q0121p0347z2p976_low}} 
\begin{tabular}{r r r r r r r r r r r}
\hline
z & $b$ &log N($\rm Fe~{\sc ii}$) &log N($\rm Si~{\sc ii}$) &log N($\rm C~{\sc ii}$) &log N($\rm O~{\sc i}$) &log N($\rm Al~{\sc iii}$) \\
\hline
2.976613& 9.90$\pm$0.94&13.18$\pm$0.30&$<$13.15 &$<$13.72 &$<$13.98&11.93$\pm$0.32 \\
\hline 				       			 	  
\end{tabular}			       			 	  
\end{center}			       			 	  
\end{table*}

\begin{figure}
\begin{center}
\includegraphics[height=6cm, width=8cm, angle=0]{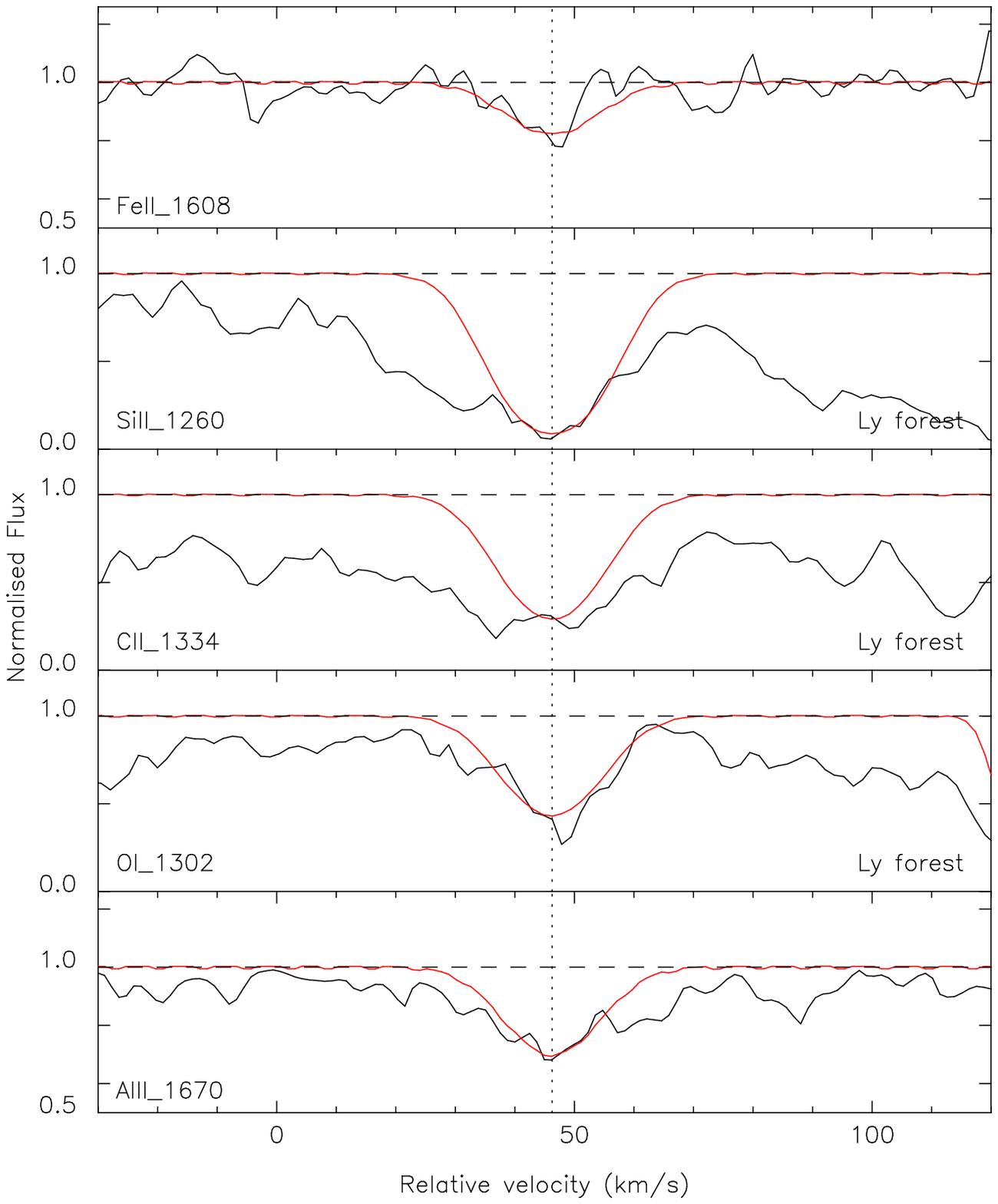}
\caption{Fit to the low-ionisation transitions of the 
\zabs=2.976 \loghi=19.53$\pm$0.10 sub-DLA towards PSS J0121$+$0347 (see Table~\ref{t:Q0121p0347z2p976_low}). 
\label{f:Q0121p0347z2p976_low}}
\end{center}
\end{figure}

\begin{table}
\begin{center}
\caption{Parameters fit to the high-ionisation transitions of the 
\zabs=2.976 \loghi=19.53$\pm$0.10 sub-DLA towards PSS J0121$+$0347. 
\label{t:Q0121p0347z2p976_high}}
\begin{tabular}{r r r r }
\hline
z & $b$ &log N($\rm C~{\sc iv}$) &log N($\rm Si~{\sc iv}$) \\
\hline
2.975965&   11.90$\pm$0.08&$<$13.50&    13.10$\pm$0.10\\
2.976579&   16.90$\pm$0.08&$<$13.47&    12.95$\pm$0.10\\
\hline 				       			 	  
\end{tabular}			       			 	  
\end{center}			       			 	  
\end{table}

\begin{figure}
\begin{center}
\includegraphics[height=8cm, width=6cm, angle=-90]{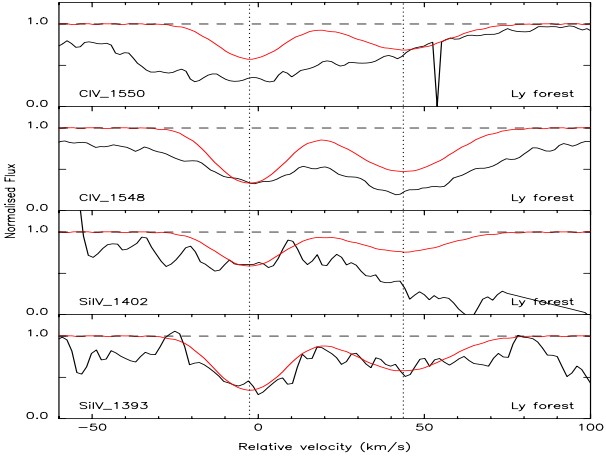}
\caption{Fit to the high-ionisation transitions of the 
\zabs=2.976 \loghi=19.53$\pm$0.10 sub-DLA towards PSS J0121$+$0347 (see Table~\ref{t:Q0121p0347z2p976_high}). 
\label{f:Q0121p0347z2p976_high}}
\end{center}
\end{figure}

%%%%%%%%%%%%%%%%%%%%%%%%%%%%%%%%%%%%%%%%%%%%%%%%%%%%%%%%%%%%%%%%%%%%%%%%%%%%%%%%%%%%%%%%%%%%%%%%%

\vspace{0.5cm}
\item{{\bf SDSS J0124$+$0044 (\zem=3.840, \zabs=2.988, \loghi=19.18$\pm$0.10):} 

This system has low-ionisation lines which cover a rather large
velocity range and the profiles are well characterised by two separate
clumps at $\sim-$100 km~s$^{-1}$ and $\sim+$30 km~s$^{-1}$.  A
simultaneous fit to \alii 1670 and \siii 1526 is used to derive the
redshifts and $b$ of the 4 components in the low-ionisation
transitions of the sub-DLA at \zabs=2.988 towards SDSS
J0124$+$0400. These results are checked for consistency with many other
$\rm Si~{\sc iii}$ lines (\siii 1193, \siii 1190, \siii 1260 and \siii 1526 -- \siii 1808
is not covered by our data). The same redshifts and $b$ are used to
derive the column density of \cii 1334 and upper limits
from \sii 1253, \sii 1259 and \feiii 1122 which are all probably blended
by lines from the Lyman-$\alpha$ forest. Similarly, an upper limit is
derived for $\rm Fe~{\sc ii}$ using the many lines falling in the forest (\feii
1063, \feii 1096, \feii 1121, \feii 1125, \feii 1143 and \feii 1144) as
well as \feii 1608. The fit is shown in
Figure~\ref{f:Q0124+0044z2p988_low} and the matching parameters are
presented in Table~\ref{t:Q0124+0044z2p988_low}. The two $\rm Al~{\sc iii}$ lines
are not covered by our spectrum for this system.

The $\rm C~{\sc iv}$ doublet is nicely fitted with 7 components, the two first
being the strongest ones. The $\rm Si~{\sc iv}$ doublet is blended but is
nevertheless fitted using the same components in order to derive an
upper limit on the abundance. Note however, that there is no physical
reason suggesting that $\rm C~{\sc iv}$ and $\rm Si~{\sc iv}$ should come from the same region,
i.e. should have the same velocity profile. It is just an empirical
fact that in DLAs, this is often the case. The fit is shown in
Figure~\ref{f:Q0124+0044z2p988_high} and the matching parameters are
presented in Table~\ref{t:Q0124+0044z2p988_high}. 

}

\begin{table*}
\begin{center}
\caption{Parameters fit to the low-ionisation transitions of the 
\zabs=2.988 \loghi=19.18$\pm$0.10 sub-DLA towards SDSS J0124$+$0044. 
\label{t:Q0124+0044z2p988_low}} 
\begin{tabular}{r r r r r r r r r r r}
\hline
z      & $b$    &       log N($\rm Fe~{\sc ii}$)&log N($\rm Si~{\sc ii}$)&      log N($\rm C~{\sc ii}$)&     log N($\rm S~{\sc ii}$)&      log N($\rm Al~{\sc ii}$)&       log N($\rm Fe~{\sc iii}$)\\     
\hline
2.986404& 2.90$\pm$0.01&    $<$12.62&13.41$\pm$0.13&    $<$15.67&    $<$13.48&    11.68$\pm$0.16&    $<$13.39\\
2.986680&12.50$\pm$0.61&    $<$13.20&13.98$\pm$0.11&    $<$14.62&    $<$13.59&    12.58$\pm$0.12&    $<$13.59\\
2.988448&12.80$\pm$0.03&    $<$13.00&12.71$\pm$0.16&    $<$13.53&    $<$13.88&    11.82$\pm$0.13&    $<$13.51\\
2.988998& 8.50$\pm$0.20&    $<$12.72&12.78$\pm$0.16&    $<$14.07&    $<$13.63&    11.93$\pm$0.14&    $<$13.44\\
\hline
\end{tabular}			       			 	  
\end{center}			       			 	  
\end{table*}

\begin{figure}
\begin{center}
\includegraphics[height=8cm, width=12cm, angle=-90]{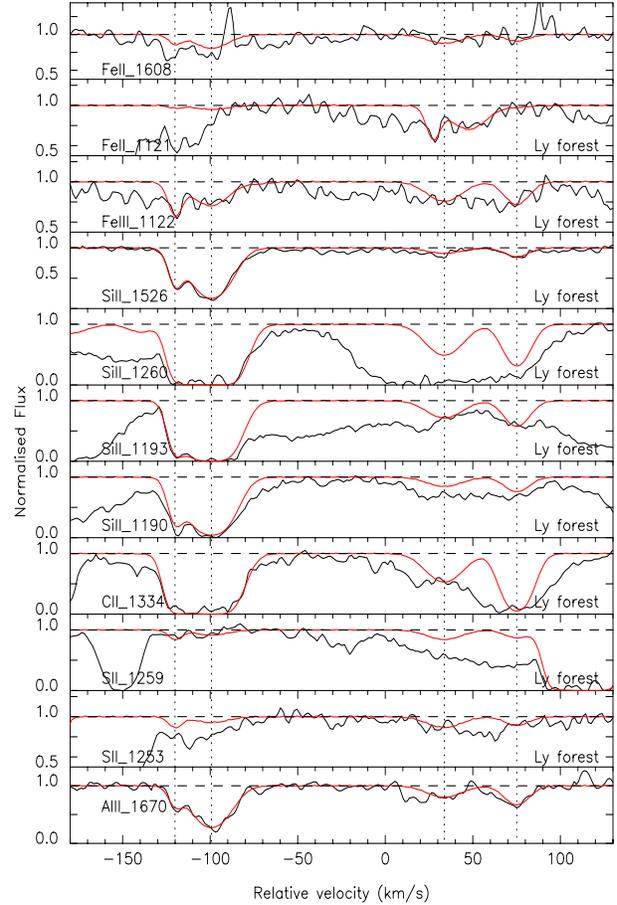}
\caption{Fit to the low-ionisation transitions of the 
\zabs=2.988 \loghi=19.18$\pm$0.10 sub-DLA towards SDSS J0124$+$0044 (see Table~\ref{t:Q0124+0044z2p988_low}). 
\label{f:Q0124+0044z2p988_low}}
\end{center}
\end{figure}

\begin{table}
\begin{center}
\caption{Parameters fit to the high-ionisation transitions of the 
\zabs=2.988 \loghi=19.18$\pm$0.10 sub-DLA towards SDSS J0124$+$0044. 
\label{t:Q0124+0044z2p988_high}}
\begin{tabular}{r r r r }
\hline
z      & $b$    &log N($\rm C~{\sc iv}$)   &log N($\rm Si~{\sc iv}$)   \\ 
\hline
2.986426&    9.40$\pm$0.01& 13.60$\pm$0.02   &    $<$13.00\\
2.986677&   14.70$\pm$0.61& 14.19$\pm$0.18   &    $<$13.59\\
2.987084&   10.70$\pm$0.03& 13.44$\pm$0.02   &    $<$13.34\\
2.987509&   15.60$\pm$0.02& 13.31$\pm$0.08   &    $<$13.41\\
2.988232&   11.94$\pm$0.10& 13.10$\pm$1.04   &    $<$13.34\\
2.988896&   17.70$\pm$0.05& 13.01$\pm$0.07   &    $<$13.47\\
2.989483&    5.70$\pm$0.05& 12.33$\pm$1.41   &    $<$12.97\\
\hline 				       			 	  
\end{tabular}			       			 	  
\end{center}			       			 	  
\end{table}

\begin{figure}
\begin{center}
\includegraphics[height=8cm, width=6cm, angle=-90]{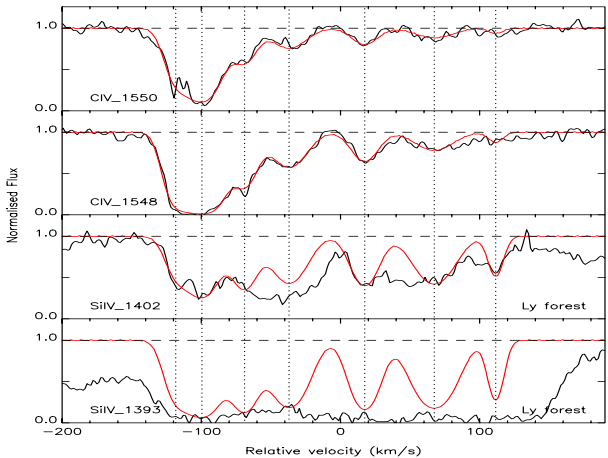}
\caption{fit to the high-ionisation transitions of the 
\zabs=2.988 \loghi=19.18$\pm$0.10 sub-DLA towards SDSS J0124$+$0044 (Table~\ref{t:Q0124+0044z2p988_high}). 
\label{f:Q0124+0044z2p988_high}} 
\end{center}
\end{figure}

%%%%%%%%%%%%%%%%%%%%%%%%%%%%%%%%%%%%%%%%%%%%%%%%%%%%%%%%%%%%%%%%%%%%%%%%%%%%%%%%%%%%%%%%%%%%%%%%%

\vspace{0.5cm}
\item{{\bf SDSS J0124$+$0044 (\zem=3.840, \zabs=3.078, \loghi=20.21$\pm$0.10):}
 
This system has fairly high \hi\ column density of
\loghi=20.21$\pm$0.10. The \siii 1304 and \siii 1526 profiles of the
sub-DLA at \zabs=3.078 are well fitted by two components. The results
of this simultaneous fit is applied on the blended \siii 1190, \siii
1193 and \siii 1260 for consistency check. The same redshifts and
$b$ are used to put limits on \oi 1302, \cii 1334 and \feii
1096, \feii 1125, \feii 1144 and \feii 1608. $\rm O~{\sc i}$ and $\rm C~{\sc ii}$ are clearly
saturated and lead to lower limits. A column density for \feii is
constrained from a combination of the red wing of \feii 1608 and the
blue wing of \feii 1096 and is consistent with all the other \feii lines
available. The fit is shown in Figure~\ref{f:Q0124+0044z3p078_low} and
the matching parameters are presented in
Table~\ref{t:Q0124+0044z3p078_low}.

The $\rm Si~{\sc iv}$ doublet associated with this system is totally blended in the
forest but the $\rm C~{\sc iv}$ seems to be detected as an extremely broad b$>$100
km~s${-1}$ single component line. The strength of the two $\rm C~{\sc iv}$ lines of
the doublet are consistent with their respective oscillator strengths,
but they do not display the characteristic complex profiles of
sub-DLA/DLAs. Therefore, care should be taken in interpreting these
features as $\rm C~{\sc iv}$. The column density we derive is N($\rm C~{\sc iv}$)=13.92$\pm$0.11
but this system might as well be a large \nhi\ system with no $\rm C~{\sc iv}$
detected. This is illustrated in Figure~\ref{f:Q0124+0044z3p078_high}.

}

\begin{table*}
\begin{center}
\caption{Parameters fit to the low-ionisation transitions of the 
\zabs=3.078 \loghi=20.21$\pm$0.10 sub-DLA towards SDSS J0124$+$0044. 
\label{t:Q0124+0044z3p078_low}} 
\label{t:fit}
\begin{tabular}{r r r r r r }
\hline
z & $b$ & log N($\rm Fe~{\sc ii}$)&log N($\rm Si~{\sc ii}$)& log N($\rm C~{\sc ii}$)& log N($\rm O~{\sc i}$)\\
\hline
3.077625&6.20$\pm$3.20&$<$13.95&13.75$\pm$0.20&$>$14.93&$>$15.00\\
3.077758&2.10$\pm$0.40&$<$13.65&15.11$\pm$0.40&$>$15.15&$>$14.60\\
\hline
\end{tabular}			       			 	  
\end{center}			       			 	  
\end{table*}

\begin{figure}
\begin{center}
\includegraphics[height=12cm, width=8cm, angle=0]{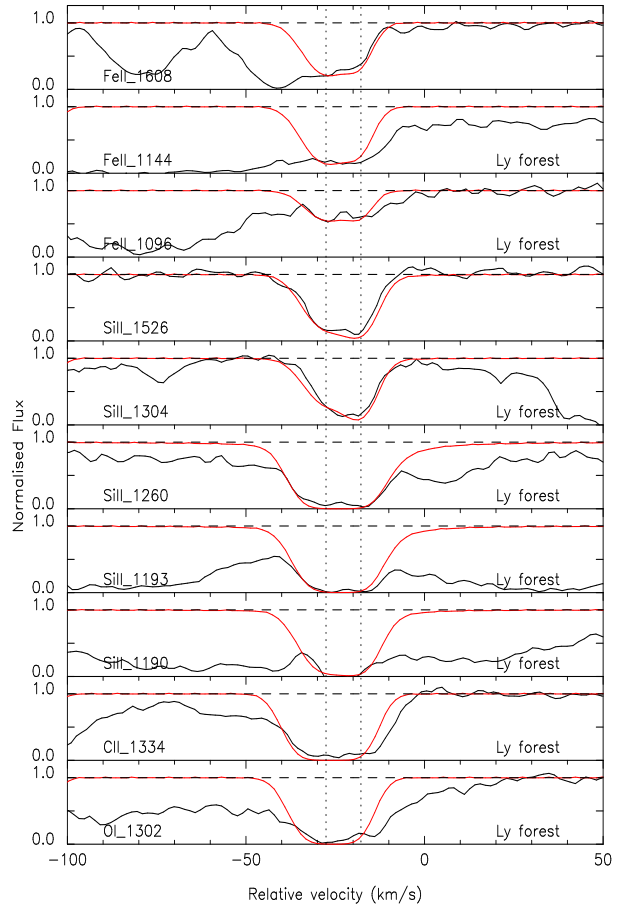}
\caption{Fit to the low-ionisation transitions of the 
\zabs=3.078 \loghi=20.21$\pm$0.10 sub-DLA towards SDSS J0124$+$0044 (see Table~\ref{t:Q0124+0044z3p078_low}). 
\label{f:Q0124+0044z3p078_low}}
\end{center}
\end{figure}

\begin{figure}
\begin{center}
\includegraphics[height=8cm, width=6cm, angle=-90]{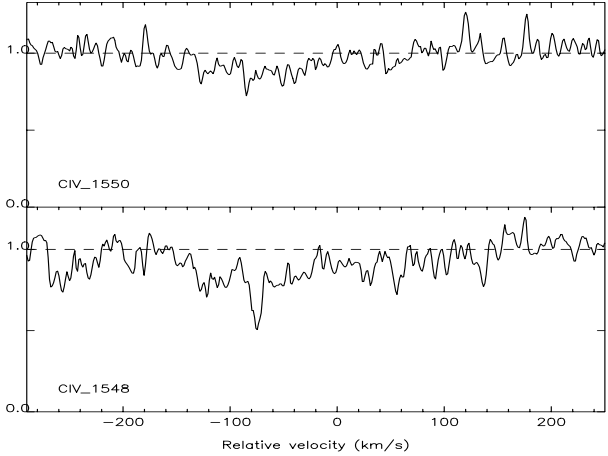}
\caption{High-ionisation ions of the absorber detected towards SDSS J0124$+$0044 with \loghi=20.21$\pm$0.10 at
\zabs=3.078.\label{f:Q0124+0044z3p078_high}} 
\end{center}
\end{figure}

%%%%%%%%%%%%%%%%%%%%%%%%%%%%%%%%%%%%%%%%%%%%%%%%%%%%%%%%%%%%%%%%%%%%%%%%%%%%%%%%%%%%%%%%%%%%%%%%%

\vspace{0.5cm}
\item{{\bf PSS J0133$+$0400 (\zem=4.154, \zabs=3.139, \loghi=19.01$\pm$0.10):}

This sub-DLA has a low \nhi\ column density with
\loghi=19.01$\pm$0.10 at \zabs=3.139. Unfortunately, many of the 
expected strong metal lines for this system fall in spectral gaps
(i.e. \feii 1608, \civ 1548, \civ 1550) or are blended with \hi\ lines from
the Lyman-$\alpha$ forest (i.e. the other \feii lines). The two
component fit was performed on \siii 1260 with a consistency check on
\siii 1808 to derive an upper limit on the total SiII column density. Attempts to increase the number of components does not
improve the $\chi^{2}$ of the fit in this case. The same redshifts and
$b$ were applied on the blended \cii 1334 to derive an upper
limit on the C abundance. The fit is shown in
Figure~\ref{f:Q0133+0400z3p139_low} and the matching parameters are
presented in Table~\ref{t:Q0133+0400z3p139_low}. \aliii 1854 is
strongly blended, but \aliii 1862 is used to derive an 4$\sigma$ upper
limit: log N($\rm Al~{\sc iii}$)$<$11.44.

The position of the high-ionisation transitions are unfortunate in  
this system too. The \siiv 1402 falls in a DLA along the same
sight-of-line while the \siiv 1393 is completely blended with \hi\
lines from the Lyman-$\alpha$ forest. The $\rm C~{\sc iv}$ doublet is situated in
one of the spectral gap of our data. }

\begin{table}
\begin{center}
\caption{Parameters fit to the low-ionisation transitions of the 
\zabs=3.139 \loghi=19.01$\pm$0.10 sub-DLA towards PSS J0133$+$0400. 
\label{t:Q0133+0400z3p139_low}} 
\begin{tabular}{r r r r }
\hline
z & $b$ &log N($\rm Si~{\sc ii}$)&log  N($\rm C~{\sc ii}$)\\
\hline
3.138360&31.60$\pm$0.20&$<$12.87&$<$13.25\\
3.139605& 7.40$\pm$0.10&$<$12.81&$<$13.72\\
\hline
\end{tabular}			       			 	  
\end{center}			       			 	  
\end{table}

\begin{figure}
\begin{center}
\includegraphics[height=8cm, width=6cm, angle=-90]{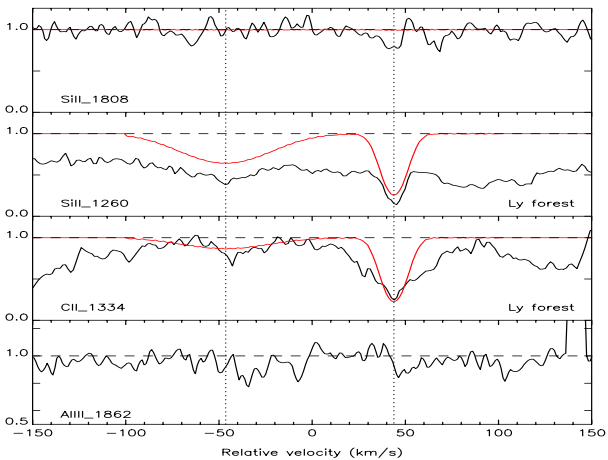}
\caption{Fit to the low-ionisation transitions of the 
\zabs=3.139 \loghi=19.01$\pm$0.10 sub-DLA towards PSS J0133$+$0400 (see Table~\ref{t:Q0133+0400z3p139_low}). The fit of \cii 1334 is used to derive an upper limit on the $\rm C~{\sc ii}$ column density. 
\label{f:Q0133+0400z3p139_low}} 
\end{center}
\end{figure}

%%%%%%%%%%%%%%%%%%%%%%%%%%%%%%%%%%%%%%%%%%%%%%%%%%%%%%%%%%%%%%%%%%%%%%%%%%%%%%%%%%%%%%%%%%%%%%%%%

\vspace{0.5cm}
\item{{\bf PSS J0133$+$0400 (\zem=4.154, \zabs=3.995, \loghi=19.94$\pm$0.15):} 

This system has a fairly high \nhi\ with \loghi=19.94$\pm$0.15 at
\zabs=3.995. The redshifts and $b$ of the 5 components were
derived from partially blended lines of \siii 1260 (telluric
contamination) for two bluest components and \siii 1190 (Lyman-$\alpha$
forest contamination) for the three remaining components. The
resulting fit was checked upon other $\rm Si~{\sc iii}$ lines (\siii 1193, \siii 1260
and \siii 1808). As a result, we can obtain a good value of the total
N($\rm Si~{\sc ii}$). Most of the \feii lines in this system are blended with
Lyman-$\alpha$ forest features (\feii 1063, \feii 1096, \feii 1121 and
\feii 1125) or fall in a zero flux gap (\feii 1143 and \feii 1144) due
to DLAs along the same line-of-sight. Only the \feii 1608 line is covered by
our spectrum but it is undetected. We derive an upper limit log
N($\rm Fe~{\sc ii}$)$<$13.56 at 4$\sigma$. Other ions are not detected and allow
the derivation of upper limits: log N($\rm Ni~{\sc ii}$)$<$12.31, log
N($\rm Al~{\sc ii}$)$<$11.36 and log N($\rm Al~{\sc iii}$)$<$11.69. The fit is shown in
Figure~\ref{f:Q0133+0400z3p995_low} and the matching parameters are
presented in Table~\ref{t:Q0133+0400z3p995_low}.

The 5 components for the high-ionisation transitions are derived from
a simultaneous fit of the lines \siiv 1393 and \siiv 1402. \siiv 1393
provides a column density determination free from any
contamination. The same redshifts and $b$ are used to fit the $\rm C~{\sc iv}$
doublet of the system and derive an upper limit on this blended
lines. The fit is shown in Figure~\ref{f:Q0133+0400z3p995_high} and
the matching parameters are presented in
Table~\ref{t:Q0133+0400z3p995_high}.  }

\begin{table}
\begin{center}
\caption{Parameters fit to the low-ionisation transitions of the 
\zabs=3.995 \loghi=19.94$\pm$0.15 sub-DLA towards PSS J0133$+$0400. 
\label{t:Q0133+0400z3p995_low}} 
\begin{tabular}{r r r }
\hline
z      & $b$    &log N($\rm Si~{\sc ii}$)\\     
\hline
3.994089& 3.10$\pm$0.16&12.24$\pm$0.12\\
3.994422& 8.50$\pm$0.11&12.83$\pm$0.11\\
3.995350& 5.90$\pm$0.21&13.56$\pm$0.14\\
3.995771&20.00$\pm$0.15&13.22$\pm$0.11\\
3.996604& 4.70$\pm$0.11&13.29$\pm$0.18\\
\hline 				       			 	  
\end{tabular}			       			 	  
\end{center}			       			 	  
\end{table}

\begin{figure}
\begin{center}
\includegraphics[height=7cm, width=8cm, angle=0]{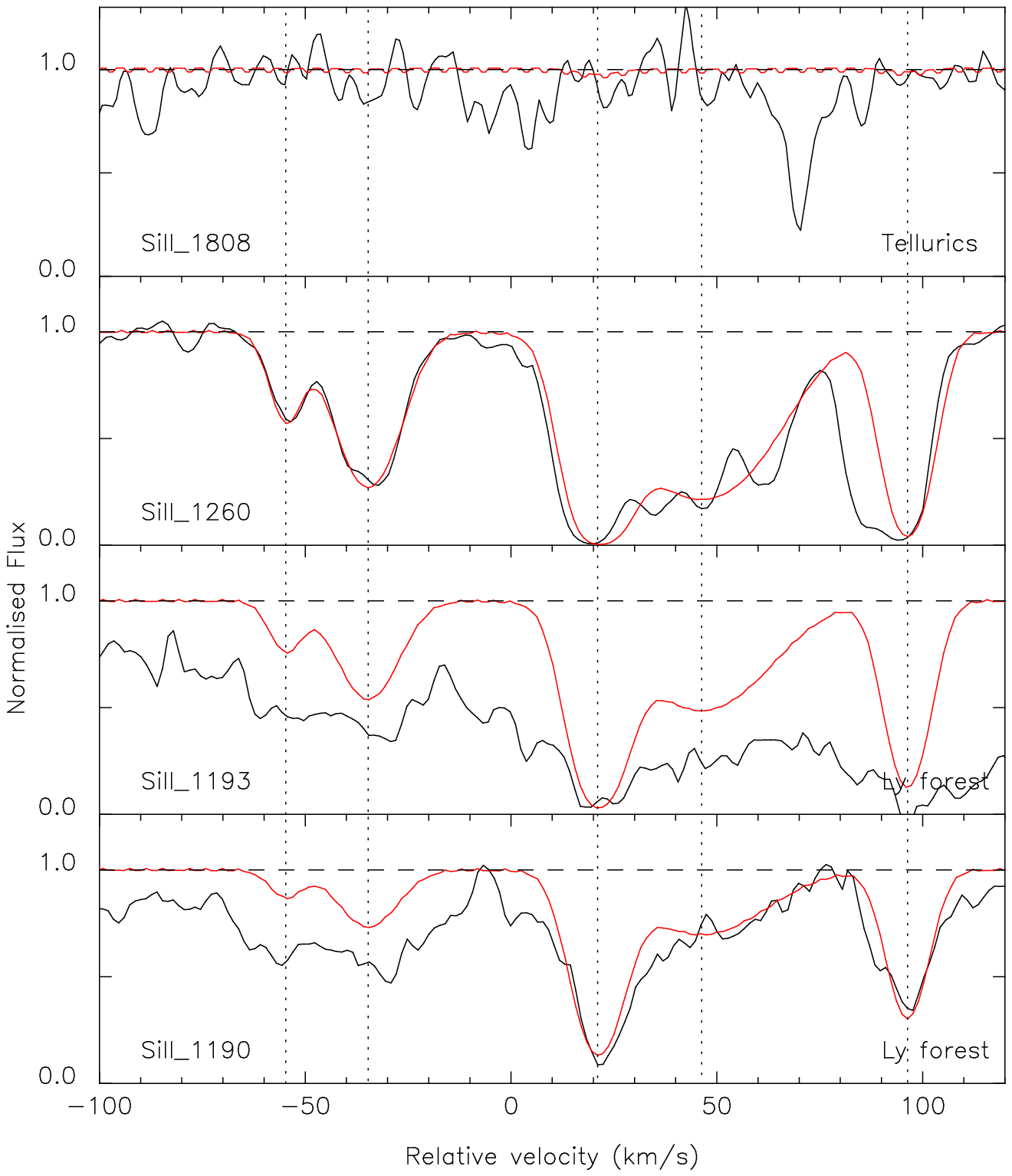}
\caption{Fit to the low-ionisation transitions of the 
\zabs=3.995 \loghi=19.94$\pm$0.15 sub-DLA towards PSS J0133$+$0400 (see Table~\ref{t:Q0133+0400z3p995_low}). 
\label{f:Q0133+0400z3p995_low}}
\end{center}
\end{figure}

\begin{table}
\begin{center}
\caption{Parameters fit to the high-ionisation transitions of the 
\zabs=3.995 \loghi=19.94$\pm$0.15 sub-DLA towards PSS J0133$+$0400. 
\label{t:Q0133+0400z3p995_high}}
\begin{tabular}{r r r r }
\hline
z      & $b$    &log N($\rm C~{\sc iv}$)   &log N($\rm Si~{\sc iv}$)     \\ 
\hline
3.994154& 8.90$\pm$ 0.01&  $<$12.92 &12.99$\pm$0.25\\
3.994482& 9.30$\pm$ 0.86&  $<$12.94 &12.90$\pm$0.03\\
3.994817& 9.30$\pm$ 0.03&  $<$12.93 &12.47$\pm$0.49\\
3.995580& 3.20$\pm$ 0.92&  $<$11.95 &12.24$\pm$0.42\\
3.996051&15.60$\pm$ 1.33&  $<$12.79 &12.89$\pm$0.01\\
\hline 				       			 	  
\end{tabular}			       			 	  
\end{center}			       			 	  
\end{table}			       			 	  

\begin{figure}
\begin{center}
\includegraphics[height=8cm, width=6cm, angle=-90]{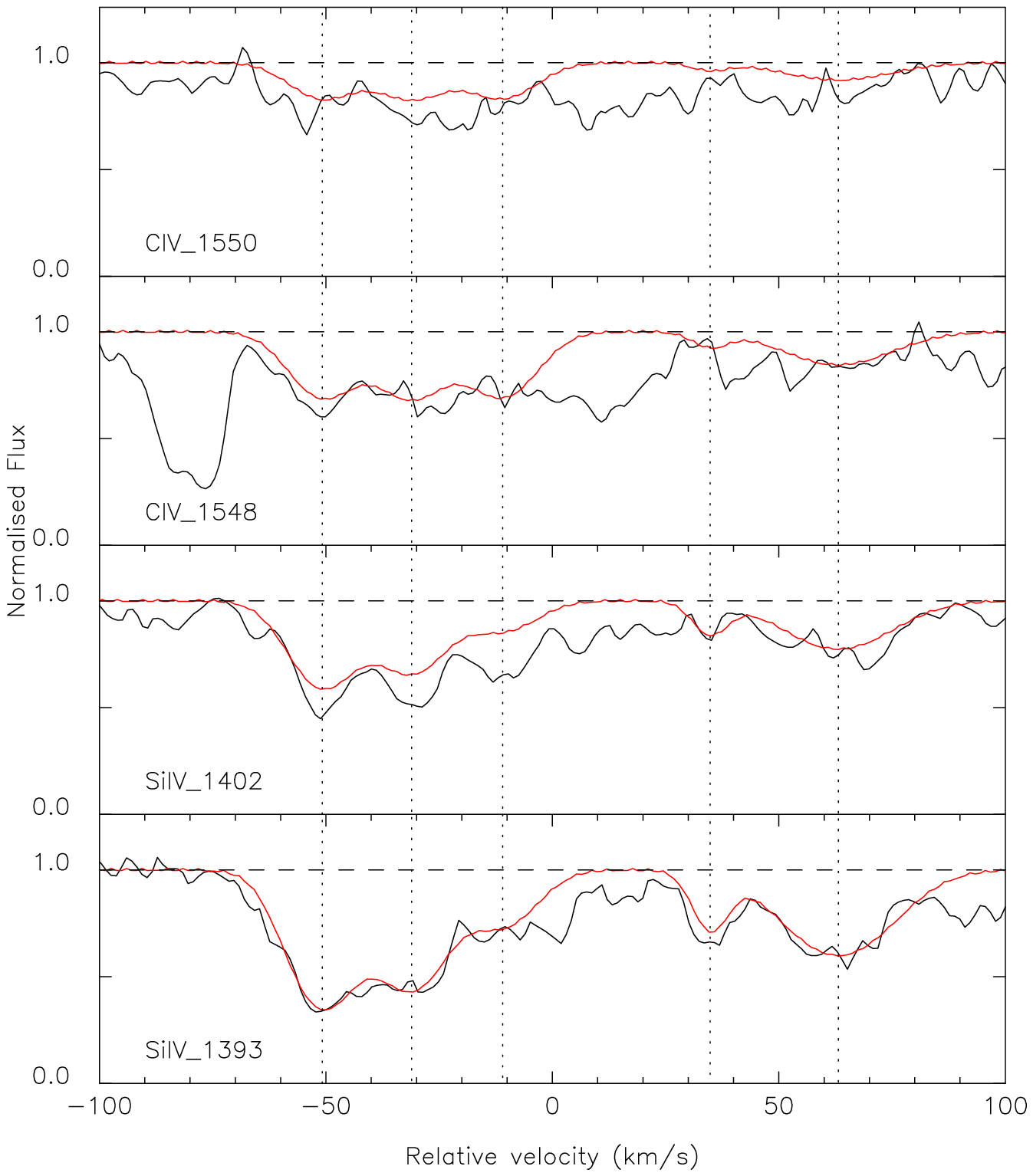}
\caption{Fit to the high-ionisation transitions of the 
\zabs=3.995 \loghi=19.94$\pm$0.15 sub-DLA towards PSS J0133$+$0400 (see Table~\ref{t:Q0133+0400z3p995_high}). 
\label{f:Q0133+0400z3p995_high}}
\end{center}
\end{figure}

%%%%%%%%%%%%%%%%%%%%%%%%%%%%%%%%%%%%%%%%%%%%%%%%%%%%%%%%%%%%%%%%%%%%%%%%%%%%%%%%%%%%%%%%%%%%%%%%%

\vspace{0.5cm}
\item{{\bf PSS J0133$+$0400 (\zem=4.154, \zabs=3.999, \loghi=19.16$\pm$0.15):}

This system is at the low end of the column density distribution for
sub-DLAs with \loghi=19.16$\pm$0.15 at
\zabs=3.999. Most $\rm Si~{\sc iii}$ lines are covered in this absorber but many are
blended with Lyman-$\alpha$ forest contamination (i.e. \siii 1190, \siii
1193) or telluric contamination (\siii 1260, \siii 1526). \siii 1304
falls in a spectral gap. Therefore, the region around the expected
position for \siii 1808 is used to derive an upper limit on the
abundance of $\rm Si~{\sc iii}$: log N($\rm Si~{\sc ii}$)$<$14.12 at 4$\sigma$. Similarly, upper
limits are derived for non-detection as follows: log N($\rm Al~{\sc ii}$)$<$11.10
from \alii 1670, log N($\rm Al~{\sc iii}$)$<$11.63 from \aliii 1854 and log
N($\rm Fe~{\sc ii}$)$<$13.56 from \feii 1608.

For the high-ionisation transitions, a fit is performed on \civ 1548
and \civ 1550 simultaneously. The same redshifts and $b$ are
used for the \siiv 1393/1402 doublet. All these lines appear broad and quite noisy
and therefore lead to upper limits. The fit is shown in
Figure~\ref{f:Q0133+0400z3p999_high} and the matching parameters are
presented in Table~\ref{t:Q0133+0400z3p999_high}.}

\begin{table}
\begin{center}
\caption{Parameters fit to the high-ionisation transitions of the 
\zabs=3.999 \loghi=19.16$\pm$0.15 sub-DLA towards PSS J0133$+$0400. 
\label{t:Q0133+0400z3p999_high}}
\begin{tabular}{r r r r }
\hline
z & $b$ &log N($\rm C~{\sc iv}$) &log N($\rm Si~{\sc iv}$) \\
\hline
3.998718&20.40$\pm$0.40&$<$13.44&$<$12.55\\
3.999390&34.60$\pm$0.30&$<$13.47&$<$13.15\\
\hline 				       			 	  
\end{tabular}			       			 	  
\end{center}			       			 	  
\end{table}			       			 	  

\begin{figure}
\begin{center}
\includegraphics[height=8cm, width=6cm, angle=-90]{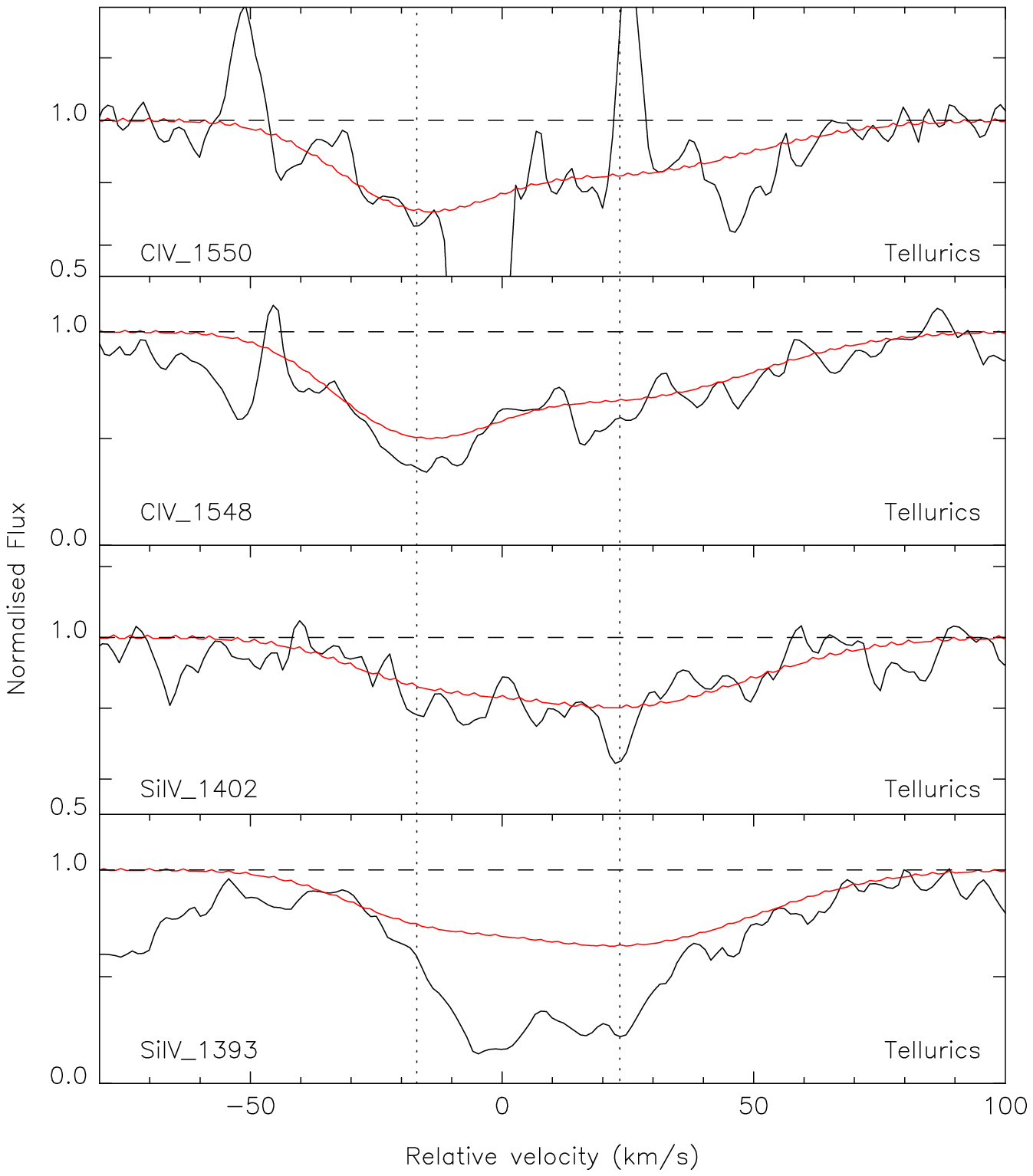}
\caption{Fit to the high-ionisation transitions of the 
\zabs=3.999 \loghi=19.16$\pm$0.15 sub-DLA towards PSS J0133$+$0400 (see Table~\ref{t:Q0133+0400z3p999_high}). 
\label{f:Q0133+0400z3p999_high}} 
\end{center}
\end{figure}

%%%%%%%%%%%%%%%%%%%%%%%%%%%%%%%%%%%%%%%%%%%%%%%%%%%%%%%%%%%%%%%%%%%%%%%%%%%%%%%%%%%%%%%%%%%%%%%%%

\vspace{0.5cm}
\item{{\bf PSS J0133$+$0400 (\zem=4.154, \zabs=4.021, \loghi=19.09$\pm$0.15):}

This is also a system at the low end of the column density range of
sub-DLAs with \loghi=19.09$\pm$0.15 at \zabs=4.021. Again, many lines
for this absorber are blended in the Lyman-$\alpha$ forest. Upper limits
from non-detections of \alii 1670, \siii 1808 and \feii 1608 are derived
at 4$\sigma$ level: log N($\rm Al~{\sc ii}$)$<$11.20, log N($\rm Si~{\sc ii}$)$<$14.05 and log
N($\rm Fe~{\sc ii}$)$<$13.56. 

A kinematically simple broad absorption is detected at the position of
the $\rm Si~{\sc iv}$ doublet. The profiles are fitted using both lines
simultaneously to derive an upper limit. The $\rm C~{\sc iv}$ doublet however is not detected in this
system. An upper limit on the abundance is derived to be log
N($\rm C~{\sc iv}$)$<$12.31. The fit is shown in
Figure~\ref{f:Q0133+0400z4p021_high} and the matching parameters are
presented in Table~\ref{t:Q0133+0400z4p021_high}.  }

\begin{table}
\begin{center}
\caption{Parameters fit to the high-ionisation transitions of the 
\zabs=4.021 \loghi=19.09$\pm$0.15 sub-DLA towards PSS J0133$+$0400. 
\label{t:Q0133+0400z4p021_high}}
\begin{tabular}{r r r }
\hline
z      & $b$     &log N($\rm Si~{\sc iv}$)  \\ 
\hline
4.021325&30.3$\pm$1.00&$<$12.89\\
\hline 				       			 	  
\end{tabular}			       			 	  
\end{center}			       			 	  
\end{table}			       			 	  

\begin{figure}
\begin{center}
\includegraphics[height=8cm, width=6cm, angle=-90]{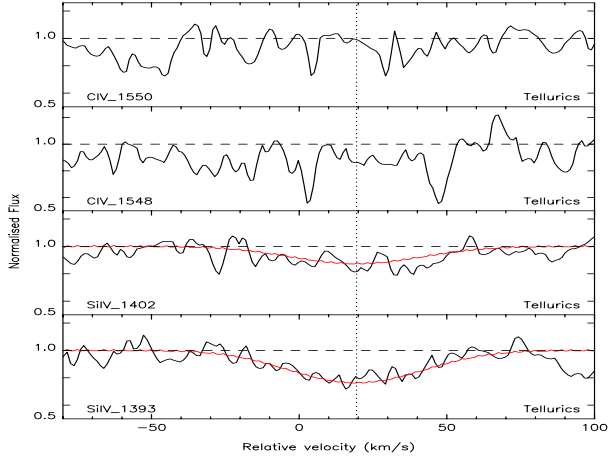}
\caption{Fit to the high-ionisation transitions of the 
\zabs=4.021 \loghi=19.09$\pm$0.15 sub-DLA towards PSS J0133$+$0400 (see Table~\ref{t:Q0133+0400z4p021_high}). 
\label{f:Q0133+0400z4p021_high}} 
\end{center}
\end{figure}

%%%%%%%%%%%%%%%%%%%%%%%%%%%%%%%%%%%%%%%%%%%%%%%%%%%%%%%%%%%%%%%%%%%%%%%%%%%%%%%%%%%%%%%%%%%%%%%%%

\vspace{0.5cm}
\item{{\bf BRI J0137$-$4224 (\zem=3.970, \zabs=3.101, \loghi=19.81$\pm$0.10):} 

The lower redshift sub-DLAs of this quasar has many metal lines
falling in the Lyman-$\alpha$ forest. With a column density
\loghi=19.81$\pm$0.10 at \zabs=3.101, it is situated in the mid-range of
the sub-DLA definition. Nevertheless, some strong lines are
undoubtfully clear from any contamination. \siii 1526 and \alii 1670 in
particular are used to derive the redshifts and $b$ of the 7
components used to fit the metal-line profile. The resulting z and $b$ are used to
derive the abundance of $\rm Fe~{\sc ii}$ from a fit to the \feii 1608 line. The
remaining \feii lines are heavily blended with Lyman-$\alpha$ forest
contamination (i.e. \feii 1096, \feii 1144). The fit is shown in
Figure~\ref{f:Q0137+4224z3p101_low} and the matching parameters are
presented in Table~\ref{t:Q0137+4224z3p101_low}.

Concerning the high-ionisation transitions, the $\rm Si~{\sc iv}$ doublet is
falling in the Lyman-$\alpha$ forest, therefore totally blended and
not useful for abundance determination. Similarly, the $\rm C~{\sc iv}$ doublet is
blended with unidentified lines as well as with apparent emission
lines which probably are the products of bad cosmic rays clipping (no
telluric lines were identified in this region) and have been removed
to perform the fit. The \civ 1550 is fitted with 8 components and the
resulting profile is made consistent with \civ 1548. 
The fit is shown in
Figure~\ref{f:Q0137+4224z3p101_high} and the matching parameters are
presented in Table~\ref{t:Q0137+4224z3p101_high}. }

\begin{table*}
\begin{center}
\caption{Parameters fit to the low-ionisation transitions of the 
\zabs=3.101 \loghi=19.81$\pm$0.10 sub-DLA towards BRI J0137$-$4224. 
\label{t:Q0137+4224z3p101_low}}
\begin{tabular}{r r r r r}
\hline
z      & $b$  &log N($\rm Fe~{\sc ii}$)  &log N($\rm Si~{\sc ii}$)&log N($\rm Al~{\sc iii}$)\\     
\hline
3.100382&   2.90$\pm$2.64& 12.65$\pm$0.39&       12.92$\pm$0.33& 11.73$\pm$ 0.14\\
3.100559&   4.90$\pm$0.11& 12.95$\pm$0.01&	13.40$\pm$0.20& 11.88$\pm$ 0.89\\
3.100821&   7.40$\pm$0.91& 12.70$\pm$0.01&	13.64$\pm$0.29& 12.20$\pm$ 0.72\\
3.101004&   3.30$\pm$0.12& 12.80$\pm$0.12&	12.73$\pm$0.22& 11.40$\pm$ 0.54\\
3.101679&   8.00$\pm$0.77& 13.20$\pm$0.01&	13.45$\pm$0.25& 12.33$\pm$ 0.49\\
3.101915&   3.20$\pm$0.22& 12.70$\pm$0.01&	13.25$\pm$0.10& 12.06$\pm$ 0.10\\
3.102078&   4.10$\pm$0.40& 12.10$\pm$0.01&	12.27$\pm$0.20& 11.48$\pm$ 0.20\\
\hline
\end{tabular}			       			 	  
\end{center}			       			 	  
\end{table*}

\begin{figure}
\begin{center}
\includegraphics[height=6.5cm, width=8cm, angle=0]{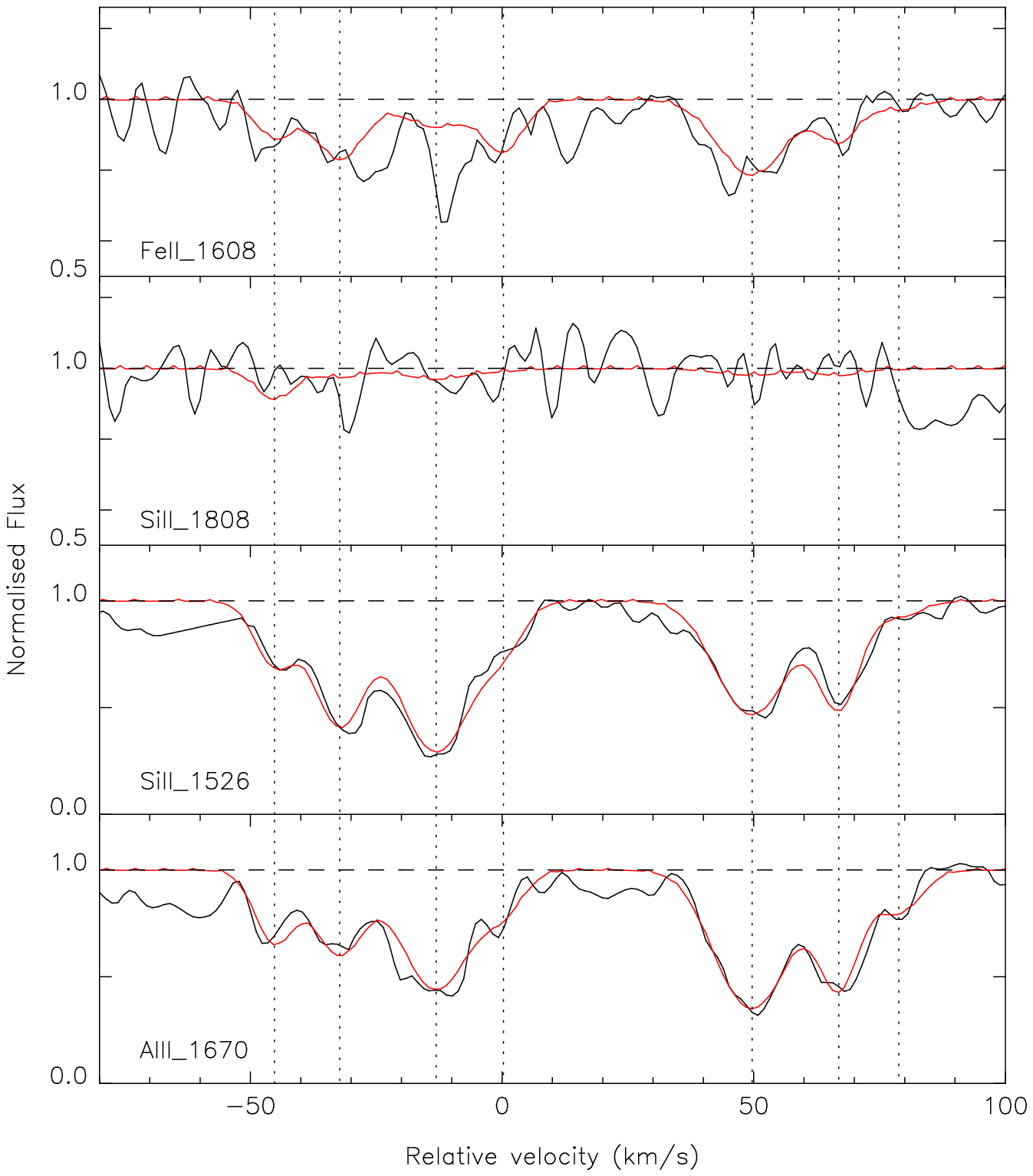}
\caption{Fit to the low-ionisation transitions of the 
\zabs=3.101 \loghi=19.81$\pm$0.10 sub-DLA towards BRI J0137$-$4224 (see Table~\ref{t:Q0137+4224z3p101_low}). 
\label{f:Q0137+4224z3p101_low}}
\end{center}
\end{figure}

\begin{table}
\begin{center}
\caption{Parameters fit to the high-ionisation transitions of the 
\zabs=3.101 \loghi=19.81$\pm$0.10 sub-DLA towards BRI J0137$-$4224. 
\label{t:Q0137+4224z3p101_high}}
\begin{tabular}{r r r  }
\hline
z      & $b$     &log N($\rm C~{\sc iv}$)     \\ 
\hline
3.100324& 3.50$\pm$0.16&  12.74$\pm$0.18\\
3.100453& 3.20$\pm$0.07&  12.43$\pm$0.10\\
3.100640& 4.50$\pm$0.28&  12.55$\pm$0.14\\
3.100995& 3.10$\pm$0.24&  12.38$\pm$0.10\\
3.101195&20.40$\pm$0.12& 12.34$\pm$0.34\\
3.101350& 3.20$\pm$0.23&  12.57$\pm$0.10\\
3.101517& 3.20$\pm$0.10&  12.79$\pm$0.25\\
3.101620& 9.90$\pm$0.20&  12.83$\pm$0.13\\
\hline				       			 	  
\end{tabular}			       			 	  
\end{center}			       			 	  
\end{table}

\begin{figure}
\begin{center}
\includegraphics[height=3cm, width=8cm, angle=0]{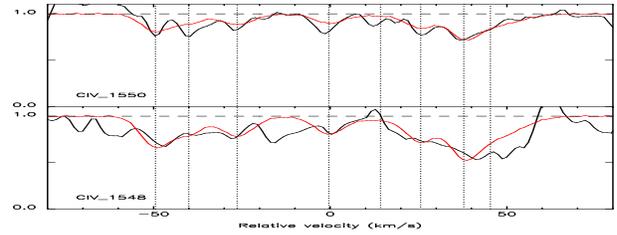}
\caption{Fit to the high-ionisation transitions of the 
\zabs=3.101 \loghi=19.81$\pm$0.10 sub-DLA towards BRI J0137$-$4224 (see Table~\ref{t:Q0137+4224z3p101_high}). 
\label{f:Q0137+4224z3p101_high}}
\end{center}
\end{figure}

%%%%%%%%%%%%%%%%%%%%%%%%%%%%%%%%%%%%%%%%%%%%%%%%%%%%%%%%%%%%%%%%%%%%%%%%%%%%%%%%%%%%%%%%%%%%%%%%%

\vspace{0.5cm}
\item{{\bf BRI J0137$-$4224 (\zem=3.970, \zabs=3.665, \loghi=19.11$\pm$0.10):} 

This sub-DLA has a fairly low column density of \loghi=19.11$\pm$0.10
at \zabs=3.665. Nevertheless, it shows strong, kinematically simple,
metal lines which are well fitted by a single component. Three lines
(\oi 1302, \siii 1260 and \cii 1334) clear from any contamination are
fitted simultaneously to derive the redshift, $b$ and
appropriate column densities. The resulting fit is overplotted on the
many blended $\rm Si~{\sc iii}$ lines (\siii 1304, \siii 1190, \siii 1190, \siii 1526
and \siii 1808) and is found to be consistent in all cases. Unluckily,
many of the \feii lines for system fall in regions contaminated by
Lyman-$\alpha$ forest lines (\feii 1063, \feii 1096, \feii 1121, \feii
1125, \feii 1143 and \feii 1144). Only \feii 1608 lies in a region free
from any contamination but the S/N of the spectrum is low. This leads
to an upper limit log N($\rm Fe~{\sc ii}$)$<$13.59 at 4$\sigma$. \alii 1670 and
\aliii 1854/\aliii 1862 are covered by our spectrum but no lines are
detected. We derive upper limits on the column densities of these ions: log
N($\rm Al~{\sc ii}$)$<$11.06 and log N($\rm Al~{\sc iii}$)$<$11.96. The fit is shown in
Figure~\ref{f:Q0137-4224z3p665_low} and the matching parameters are
presented in Table~\ref{t:Q0137-4224z3p665_low}.

The high-ionisation transitions of this system present a very
particular signature. Both lines of the $\rm Si~{\sc iv}$ doublet (1393 and 1402)
are clearly detected and very well fitted by a simple two component
profile. But rather surprisingly, both $\rm C~{\sc iv}$ lines (1548 and 1550) of
the $\rm C~{\sc iv}$ doublet are covered by our data but {\it seem undetected}. In
fact, the weak component falling at the expected redshift in \civ 1550
happens to be a telluric line, although the \civ 1548 region is free
from telluric contamination. At any rate, this line is within the
noise of the spectrum and could not be assumed to be a real line for
sure. From the S/N in the region where the doublet is expected to
fall, we deduce an upper limit on the column density log
N($\rm C~{\sc iv}$)$<$12.11 at 4$\sigma$. Interestingly in Dessauges-Zavadsky et
al. (2003), we report the detection in a sub-DLA of $\rm C~{\sc iv}$ lines with no
$\rm Si~{\sc iv}$ associated. But to our knowledge, this is the first time that a
large \nhi\ absorber is reported to have $\rm Si~{\sc iv}$ but no $\rm C~{\sc iv}$
detected. This is could be a signature of $\rm Si~{\sc iv}$ associated with the
\nhi\ gas whilst $\rm C~{\sc iv}$ will be part of an external shell surrounding this
region. The fit is shown in Figure~\ref{f:Q0137-4224z3p665_high}
and the matching parameters are presented in
Table~\ref{t:Q0137-4224z3p665_high}.}

\begin{table*}
\begin{center}
\caption{Parameters fit to the low-ionisation transitions of the 
\zabs=3.665 \loghi=19.11$\pm$0.10 sub-DLA towards BRI J0137$-$4224. 
\label{t:Q0137-4224z3p665_low}}
\begin{tabular}{r r r r r}
\hline
z      & $b$    &log N($\rm Si~{\sc ii}$)&log N($\rm C~{\sc ii}$)&log N($\rm O~{\sc i}$)\\     
\hline
3.665111&7.02$\pm$0.50&12.40$\pm$0.13&13.14$\pm$0.13&13.38$\pm$0.13\\
\hline 				       			 	  
\end{tabular}			       			 	  
\end{center}			       			 	  
\end{table*}

\begin{figure}
\begin{center}
\includegraphics[height=6cm, width=6cm, angle=-90]{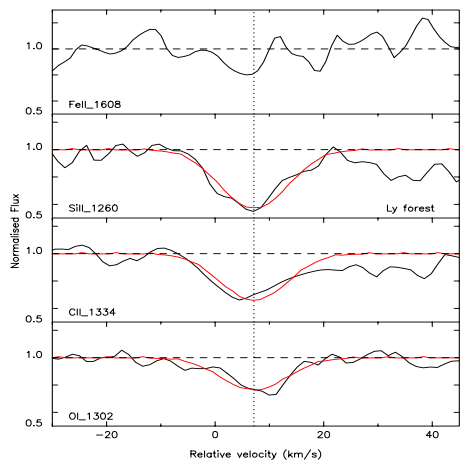}
\caption{Fit to the low-ionisation transitions of the 
\zabs=3.665 \loghi=19.11$\pm$0.10 sub-DLA towards BRI J0137$-$4224 (see Table~\ref{t:Q0137-4224z3p665_low}). The region 25--25 
km s$^{-1}$ is contaminated by an unidentified blend.
\label{f:Q0137-4224z3p665_low}}
\end{center}
\end{figure}

\begin{table}
\begin{center}
\caption{Parameters fit to the high-ionisation transitions of the 
\zabs=3.665 \loghi=19.11$\pm$0.10 sub-DLA towards BRI J0137$-$4224. 
\label{t:Q0137-4224z3p665_high}}
\begin{tabular}{r r r }
\hline
z & $b$ &log N($\rm Si~{\sc iv}$) \\
\hline
3.665009&6.60$\pm$5.20&12.43$\pm$0.22\\
3.665482&7.30$\pm$3.80&12.95$\pm$0.10\\
\hline
\end{tabular}			       			 	  
\end{center}			       			 	  
\end{table}

\begin{figure}
\begin{center}
\includegraphics[height=8cm, width=6cm, angle=-90]{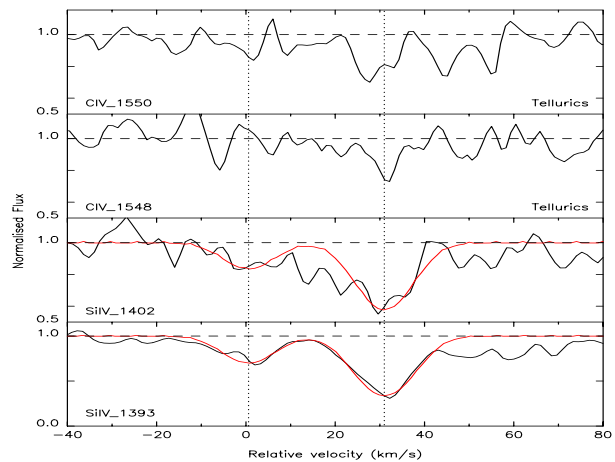}
\caption{Fit to the high-ionisation transitions of the 
\zabs=3.665 \loghi=19.11$\pm$0.10 sub-DLA towards BRI J0137$-$4224 (see Table~\ref{t:Q0137-4224z3p665_high}). 
\label{f:Q0137-4224z3p665_high}}
\end{center}
\end{figure}

%%%%%%%%%%%%%%%%%%%%%%%%%%%%%%%%%%%%%%%%%%%%%%%%%%%%%%%%%%%%%%%%%%%%%%%%%%%%%%%%%%%%%%%%%%%%%%%%%

\vspace{0.5cm}
\item{{\bf BR J2215$-$1611 (\zem=3.990, \zabs=3.656, \loghi=19.01$\pm$0.15):} 

The two sub-DLAs detected towards BR J2215$-$1611 are essentially
blended together and only come apart at the Lyman-$\gamma$ level. The
first system at \zabs=3.656 is also the weakest \hi\ column density
with \loghi=19.01$\pm$0.15. Metal lines are seldom detected partly
because of contamination with the Lyman-$\alpha$ forest lines. Few
lines (\siii 1304, \siii 1526, \siii 1808 and \feii 1608) fall
in regions free from any contamination and yet remain undetected. We
derive upper limits on the column densities of the corresponding ions:
log N($\rm Si~{\sc ii}$)$<$14.41, log N($\rm Fe~{\sc ii}$)$<$13.51 and 
log N($\rm S~{\sc ii}$)$<$13.68. The
intermediate ionisation transitions \aliii 1854 and \aliii 1862 are
falling in the region polluted by telluric lines which prevent us from
any detection or even determination of an upper limit.

The high-ionisation transitions, \civ 1548 and \civ 1550, are detected
in this system. To the immediate blue wavelength of the \civ 1548 line
lies a telluric line. The profile is well fitted by one component
whose characteristics are derived from a simultaneous fit of the two
members of the doublet, but only an upper limit is derived to be conservative. The $\rm Si~{\sc iv}$ in this sub-DLA is not detected
although both lines are covered by our data. This leads to an upper
limit on the column density: log N($\rm Si~{\sc iv}$)$<$11.69. The fit is shown in
Figure~\ref{f:Q2215-1611z3p656_high} and the matching parameters are
presented in Table~\ref{t:Q2215-1611z3p656_high}.

}

\begin{table}
\begin{center}
\caption{Parameters fit to the high-ionisation transitions of the 
\zabs=3.656 \loghi=19.01$\pm$0.15 sub-DLA towards BR J2215$-$1611. 
\label{t:Q2215-1611z3p656_high}}
\begin{tabular}{r r r  }
\hline
z & $b$ &log N($\rm C~{\sc iv}$) \\
\hline
3.656415&33.30$\pm$16.86&$<$13.22\\
\hline
\end{tabular}			       			 	  
\end{center}			       			 	  
\end{table}

\begin{figure}
\begin{center}
\includegraphics[height=8cm, width=6cm, angle=-90]{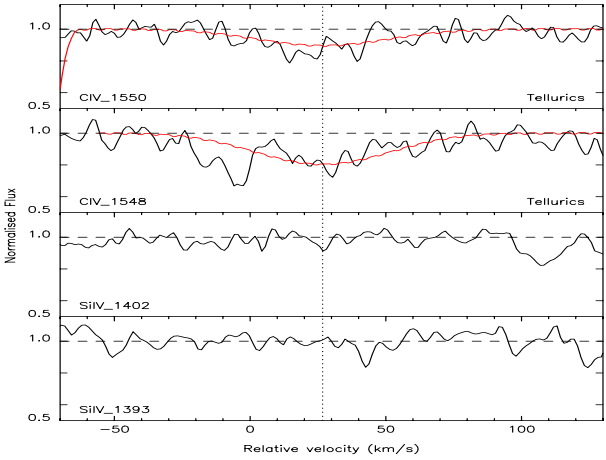}
\caption{Fit to the high-ionisation transitions of the 
\zabs=3.656 \loghi=19.01$\pm$0.15 sub-DLA towards BR J2215$-$1611 (see Table~\ref{t:Q2215-1611z3p656_high}). 
\label{f:Q2215-1611z3p656_high}} 
\end{center}
\end{figure}

%%%%%%%%%%%%%%%%%%%%%%%%%%%%%%%%%%%%%%%%%%%%%%%%%%%%%%%%%%%%%%%%%%%%%%%%%%%%%%%%%%%%%%%%%%%%%%%%%

\vspace{0.5cm}
\item{{\bf BR J2215$-$1611 (\zem=3.990, \zabs=3.662, \loghi=20.05$\pm$0.15):}

Contrary to the previous system, this sub-DLA has a fairly high \nhi\
column density \loghi=20.05$\pm$0.15 at \zabs=3.662. It is
characterised by strong metal lines showing a complex profile over
$\sim$100 km~s$^{-1}$. The 5 components are simultaneously fitted to
\oi 1302, the un-saturated components of \cii 1334, \alii 1670 and \siii
1526. The first component is narrower in \siii 1526 than in the other
metal lines including \siii 1304. We note however that \oi 1302 is controversial  
and that it is a borderline case between a lower limit and a value.

The resulting fit is applied to the other $\rm Si~{\sc ii}$ lines: \siii 1304 is
clearly well fitted by this profile. \siii 1808 is not detected in
agreement with our fit, while \siii 1190, \siii 1193 and \siii 1260 which
are situated in the Lyman-$\alpha$ forest are clearly blended by other
intervening lines. Nevertheless, the fit is found to be consistent
with the upper limit they provide. All \feii lines but \feii 1125 are
contaminated, but we use the non detection of \feii 1125 to derive log
N($\rm Fe~{\sc ii}$)$<$13.60. The fit is shown in
Figure~\ref{f:Q2215-1611z3p662_low} and the matching parameters are
presented in Table~\ref{t:Q2215-1611z3p662_low}. The intermediate
ionisation transitions \aliii 1854 and \aliii 1862 are falling in the
region polluted by telluric lines which prevent us from any detection
or even determination of an upper limit. 

The high-ionisation transitions in this system are well fitted with 3
components. The redshifts and $b$ of these were determined by
a simultaneous fit of the \civ 1548 and \siiv 1402 lines. The fit is then
superimposed on \civ 1550 and \siiv 1393. Clearly the \siiv 1393 line is
blended although no telluric lines are seen in this region. This fit
leads accurate column density determinations. The fit is shown in
Figure~\ref{f:Q2215-1611z3p662_high} and the matching parameters are
presented in Table~\ref{t:Q2215-1611z3p662_high}. }

\begin{table*}
\begin{center}
\caption{Parameters fit to the low-ionisation transitions of the 
\zabs=3.662, \loghi=20.05$\pm$0.15 sub-DLA towards BR J2215$-$1611. 
\label{t:Q2215-1611z3p662_low}}
\begin{tabular}{r r r c r r}
\hline
z & $b$ &log N($\rm Si~{\sc ii}$)&log N($\rm C~{\sc ii}$)&log N($\rm O~{\sc i}$)&log N($\rm Al~{\sc ii}$)\\
\hline
3.661105&3.30$\pm$0.10&12.95$\pm$0.28&    13.72$\pm$0.48	&    14.03$\pm$0.01&    11.70$\pm$0.16\\
3.661597&6.80$\pm$0.10&13.45$\pm$0.02&    $>$13.95		&    14.61$\pm$0.20&    11.92$\pm$0.27\\
3.661754&6.80$\pm$0.20&12.60$\pm$0.03&    $>$13.35		&    13.61$\pm$0.02&    11.47$\pm$0.10\\
3.661885&5.10$\pm$0.10&13.38$\pm$0.01&    14.81$\pm$0.20	&    14.25$\pm$0.01&    11.95$\pm$0.67\\
3.662193&3.50$\pm$0.10&13.11$\pm$0.02&    $>$14.16		&    14.58$\pm$0.02&    11.55$\pm$0.53\\
\hline 				       			 	  
\end{tabular}			       			 	  
\end{center}			       			 	  
\end{table*}

\begin{figure}
\begin{center}
\includegraphics[height=8cm, width=7cm, angle=-90]{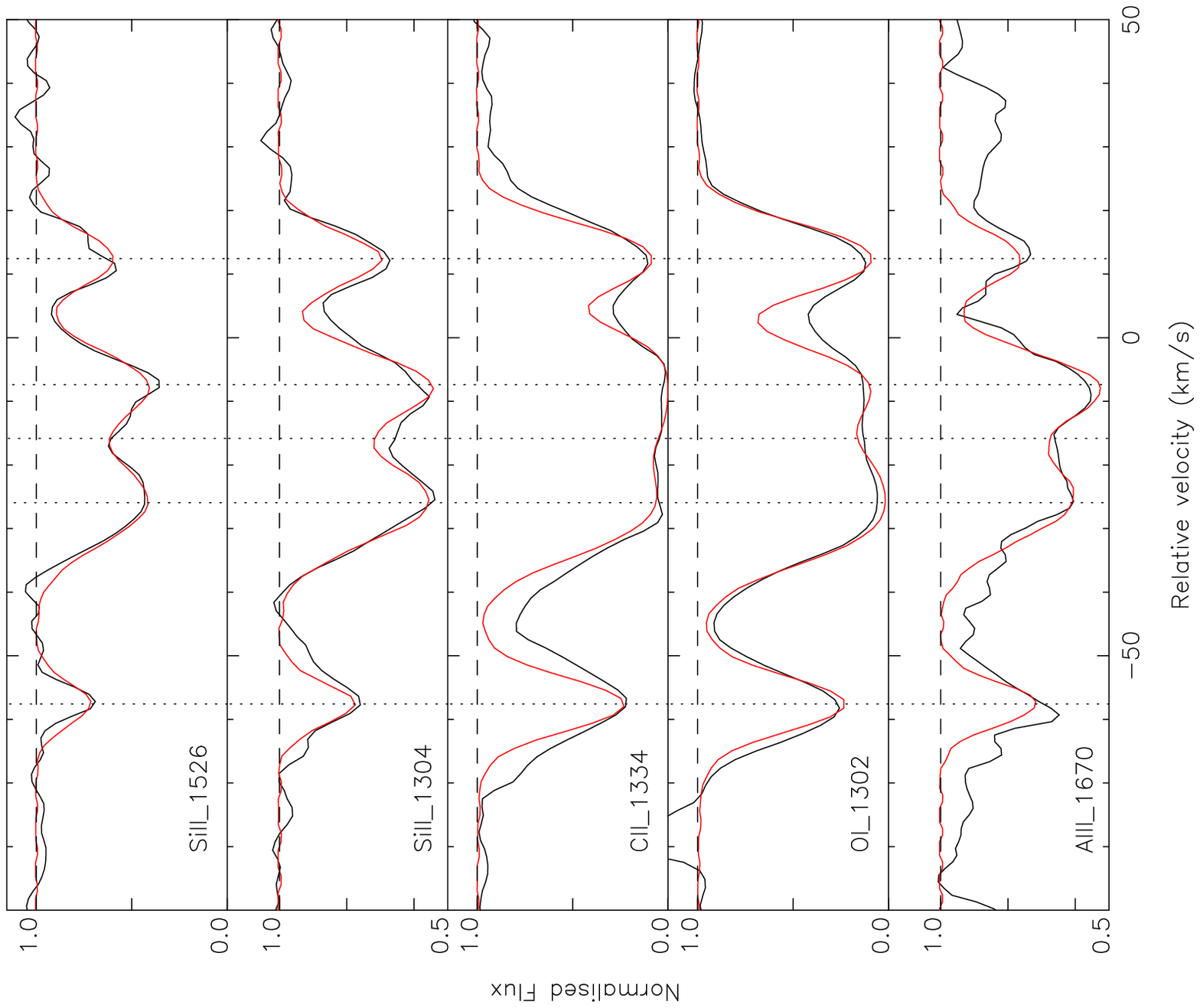}
\caption{Fit to the low-ionisation transitions of the 
\zabs=3.662, \loghi=20.05$\pm$0.15 sub-DLA towards BR J2215$-$1611 (see Table~\ref{t:Q2215-1611z3p662_low}). 
\label{f:Q2215-1611z3p662_low}} 
\end{center}
\end{figure}

\begin{table}
\begin{center}
\caption{Parameters fit to the high-ionisation transitions of the 
\zabs=3.662, \loghi=20.05$\pm$0.15 sub-DLA towards BR J2215$-$1611. 
\label{t:Q2215-1611z3p662_high}}
\label{t:fit}
\begin{tabular}{r r r r }
\hline
z      & $b$     &log N($\rm C~{\sc iv}$) &log N($\rm Si~{\sc iv}$)     \\ 
\hline
3.661409&   7.20$\pm$ 5.00&12.79$\pm$0.07&12.67$\pm$0.08\\  
3.661881&  12.40$\pm$ 3.10&13.57$\pm$0.25&13.39$\pm$0.04\\
3.662255&  26.50$\pm$ 8.80&12.92$\pm$0.08&12.53$\pm$0.07\\
\hline 				       			 	  
\end{tabular}			       			 	  
\end{center}			       			 	  
\end{table}

\begin{figure}
\begin{center}
\includegraphics[height=8cm, width=6cm, angle=-90]{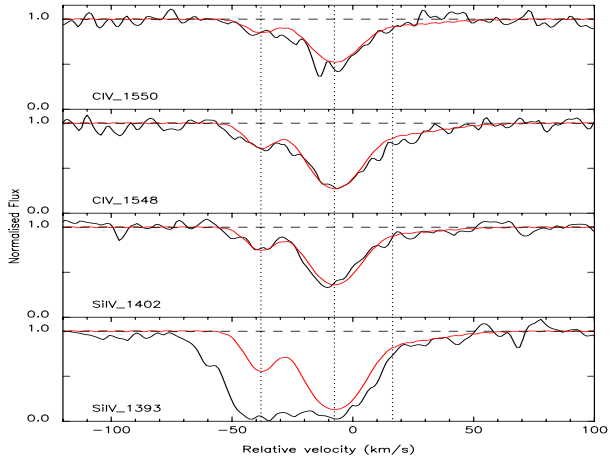}
\caption{Fit to the high-ionisation transitions of the 
\zabs=3.662, \loghi=20.05$\pm$0.15 sub-DLA towards BR J2215$-$1611 (see Table~\ref{t:Q2215-1611z3p662_high}). 
\label{f:Q2215-1611z3p662_high}} 
\end{center}
\end{figure}

%%%%%%%%%%%%%%%%%%%%%%%%%%%%%%%%%%%%%%%%%%%%%%%%%%%%%%%%%%%%%%%%%%%%%%%%%%%%%%%%%%%%%%%%%%%%%%%%%

\vspace{0.5cm}
\item{{\bf BR J2216$-$6714 (\zem=4.469, \zabs=3.368, \loghi=19.80$\pm$0.10):}

This sub-DLA has a mid-range column density \loghi=19.80$\pm$0.10 at
\zabs=3.368. Nevertheless, the low-ionisation transitions in this
system are barely detected, partly because of Lyman-$\alpha$
contamination (the metal lines with $\lambda_{\rm rest}<$1522 \AA\ in
this sub-DLA will fall in the forest). This applies to many of the
$\rm Si~{\sc iii}$ lines (\siii 1304, \siii 1260, \siii 1193, \siii 1190 and \siii 1526).
The non-detection of \siii 1808 lead to the determination of an upper
limit: log N($\rm Si~{\sc ii}$)$<$13.88 at 4$\sigma$. Similarly, most of the $\rm Fe~{\sc ii}$
lines are in the forest, but the region at the expected position of
\feii 1608 has a fairly high signal-to-noise leading to a constraining
upper limit on the $\rm Fe~{\sc ii}$ column density: log N($\rm Fe~{\sc ii}$)$<$13.26. \aliii
1854 and 1862 are also undetected: log N($\rm Al~{\sc iii}$)$<$11.51. On the
contrary, a two component feature at the expected position of \alii
1670 is clearly detected. We compute an abundance determination for
this element. The fit is shown in Figure~\ref{f:Q2216-6714z3p368_low}
and the matching parameters are presented in
Table~\ref{t:Q2216-6714z3p368_low}.

In addition, the high-ionisation transitions are also undoubtfully
detected for this system. The $\rm C~{\sc iv}$ doublet is clearly detected and
leads an accurate column density determination using a simultaneous
fit of both members of the doublet to determine the redshifts and
$b$ of 6 components. The $\rm Si~{\sc iv}$ doublet appears to be blended
with Lyman-$\alpha$ forest interlopers, although \siiv 1393 is most
probably free form any contamination. The parameters derived from $\rm C~{\sc iv}$
are used to determine the upper limits on \siiv 1393, but the first and fourth
components are not detected in  $\rm Si~{\sc iv}$. The fit is shown in
Figure~\ref{f:Q2216-6714z3p368_high} and the matching parameters are
presented in Table~\ref{t:Q2216-6714z3p368_high}.  }

\end{enumerate}

\begin{table}
\begin{center}
\caption{Parameters fit to the low-ionisation transitions of the 
\zabs=3.368, \loghi=19.80$\pm$0.10 sub-DLA towards BR J2216$-$6714. 
\label{t:Q2216-6714z3p368_low}}
\begin{tabular}{r r r }
\hline
z      & $b$    &log N($\rm Al~{\sc ii}$)\\     
\hline
3.369173&15.20$\pm$2.54&   11.88$\pm$0.26\\   
3.370053&17.40$\pm$3.04&   11.85$\pm$0.27\\   
\hline
\end{tabular}			       			 	  
\end{center}			       			 	  
\end{table}

\begin{figure}
\begin{center}
\includegraphics[height=8cm, width=1.5cm, angle=-90]{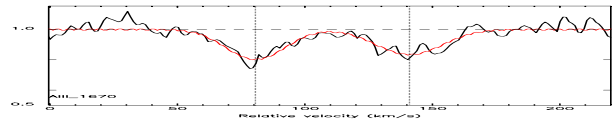}
\caption{Fit to the low-ionisation transitions of the 
\zabs=3.368, \loghi=19.80$\pm$0.10 sub-DLA towards BR J2216$-$6714 (see Table~\ref{t:Q2216-6714z3p368_low}). 
\label{f:Q2216-6714z3p368_low}}
\end{center}
\end{figure}

\begin{table}
\begin{center}
\caption{Parameters fit to the high-ionisation transitions of the 
\zabs=3.368, \loghi=19.80$\pm$0.10 sub-DLA towards BR J2216$-$6714. 
\label{t:Q2216-6714z3p368_high}}
\begin{tabular}{r r r r }
\hline
z & $b$ &log N($\rm C~{\sc iv}$) &log N($\rm Si~{\sc iv}$) \\
\hline
3.368701&  20.10$\pm$ 4.07&12.95$\pm$0.11& ...$^*$\\
3.369287&  15.50$\pm$ 2.96&13.02$\pm$0.12&$<$12.52\\
3.369919&  12.60$\pm$ 5.42&13.01$\pm$0.09&$<$12.34\\
3.370177&   6.10$\pm$ 1.99&12.58$\pm$0.22 & ...$^*$\\
3.370583&  74.00$\pm$ 1.59&13.49$\pm$0.08&$<$13.09\\
3.370990&   8.30$\pm$ 2.01&13.03$\pm$0.08 &$<$12.48\\
\hline 				       			 	  
\end{tabular}			       			 	  
\vspace{0.2cm}
\begin{minipage}{80mm}

{\bf $^*$:} These components are undetected in $\rm Si~{\sc iv}$.\\

\end{minipage}
\end{center}			       			 	  
\end{table}

\begin{figure}
\begin{center}
\includegraphics[height=8cm, width=6cm, angle=-90]{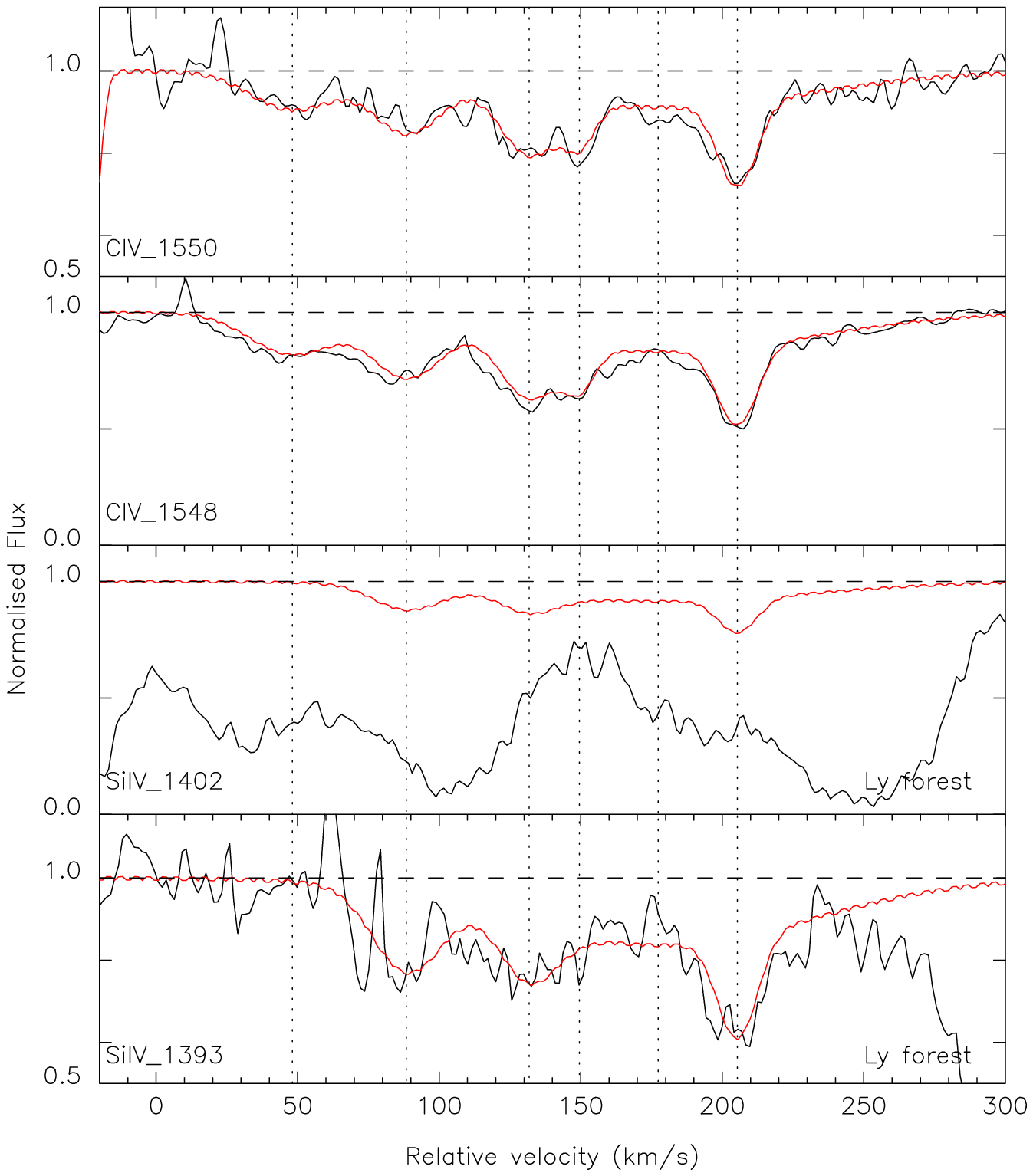}
\caption{Fit to the high-ionisation transitions of the 
\zabs=3.368, \loghi=19.80$\pm$0.10 sub-DLA towards BR J2216$-$6714 (see Table~\ref{t:Q2216-6714z3p368_high}). 
\label{f:Q2216-6714z3p368_high}}
\end{center}
\end{figure}

%%%%%%%%%%%%%%%%%%%%%%%%%%%%%%%%%%%%%%%%%%%%%%%%%%%%%%%%%%%%%%%%%%%%%%%%
%%%%%%%%%%%%%%%%%%%%%%%%%%%%%%%%%%%%%%%%%%%%%%%%%%%%%%%%%%%%%%%%%%%%%%%%

\section{Results}

\subsection{Individual Abundances}

The resulting total column densities for each elements
detected in the 13 sub-DLAs studied here are summarised in
Table~\ref{t:col} together with the error estimates.

\begin{table*}
\caption{Total column densities for each absorber. 
\label{t:col}}
\begin{tabular}{lccccccccc}
\hline
Quasar &\zabs &log N($\rm Fe~{\sc ii}$) &log N($\rm Si~{\sc ii}$) &log N($\rm C~{\sc ii}$) &log N($\rm O~{\sc i}$) &log N($\rm Al~{\sc ii}$) &log N($\rm Al~{\sc iii}$)
&log N($\rm C~{\sc iv}$) &log N($\rm Si~{\sc iv}$)\\
\hline
PSS J0118$+$0320$^\dag$ &4.128  &14.16$\pm$0.11 &14.84$\pm$0.13 &$>$17.30 &$>$16.40 &...            &...          &14.30$\pm$0.12 &13.78$\pm$0.12\\
PSS J0121$+$0347        &2.976  &13.18$\pm$0.30 &$<$13.15       &$<$13.72 &$<$13.98 &11.93$\pm$0.32 &$<$11.77$^*$ &$<$13.79       &13.33$\pm$0.10\\
SDSS J0124$+$0044$^\dag$&2.988  &$<$13.55$^+$   &14.12$\pm$0.12 &$<$15.72 &...      &12.76$\pm$0.13 &...          &14.43$\pm$0.30 &$<$14.20\\
...                     &3.078  &$<$14.13       &15.13$\pm$0.39 &$>$15.35 &$>$15.15 &...            &...          &...            &...\\
PSS J0133$+$0400        &3.139	 &...            &$<$13.14 &$<$13.85 &...      &...            &$<$11.44$^*$ &...            &...\\
...                     &3.995	 &$<$13.56$^{*@}$&13.91$\pm$0.13 &...      &...      &$<$11.36$^*$   &$<$11.69$^*$ &$<$13.51       &13.48$\pm$0.28\\
...                     &3.999  &$<$13.56$^*$   &$<$14.12$^*$   &...      &...      &$<$11.10$^*$   &$<$11.63$^*$ &$<$13.76       &$<$13.25\\
...                     &4.021	 &$<$13.56$^*$   &$<$14.05$^*$   &...      &...      &$<$11.20$^*$   &...          &$<$12.31$^*$   &$<$12.89 \\
BRI J0137$-$4224        &3.101	 &13.67$\pm$0.11 &14.11$\pm$0.25 &...      &...      &12.83$\pm$0.42 &...          &13.52$\pm$0.17 &... \\
...                     &3.665	 &$<$13.59$^*$   &12.40$\pm$0.13 &13.14$\pm$0.13&13.38$\pm$0.13 &$<$11.06$^*$   &$<$11.96$^*$ &$<$12.11$^*$   &13.06$\pm$0.13\\
BR J2215$-$1611$^\dag$	&3.656  &$<$13.51$^*$   &$<$14.41$^*$   &...      &...      &...            &...          &$<$13.22 &$<$11.69$^*$\\
...                     &3.662  &$<$13.60$^*$   &13.89$\pm$0.06 &$>$14.98 &15.05$\pm$0.02&12.46$\pm$0.45 &...          &13.71$\pm$0.21 &13.51$\pm$0.05\\
BR J2216$-$6714	        &3.368  &$<$13.26$^*$   &$<$13.88$^*$   &...      &...      &12.17$\pm$0.26 &$<$11.51$^*$ &13.88$\pm$0.10 &$<$13.32\\
\hline			
\end{tabular}  
\vspace{0.2cm}
\begin{minipage}{160mm}

{\bf $^*$:} 4 $\sigma$ upper limits corresponding to non-detections\\
{\bf $^+$:} log N($\rm Fe~{\sc iii}$)$<$14.09 in this system\\
{\bf $^@$:} log N($\rm Ni~{\sc ii}$)$<$12.31 in this system from a non-detection\\
{\bf $^\dag$:} PSS J0118$+$0320, \zabs=4.128 has log N($\rm S~{\sc ii}$)=14.26$\pm$0.11, 
SDSS J0124$+$0044, \zabs=2.988 has log N($\rm S~{\sc ii}$)$<$14.27, BR J2215$-$1611, \zabs=3.656 has log N($\rm S~{\sc ii}$)$<$13.68 from a non-detection.\\

\end{minipage}
\end{table*}

\begin{figure}
\begin{center}
\includegraphics[height=7cm, width=7cm, angle=-90]{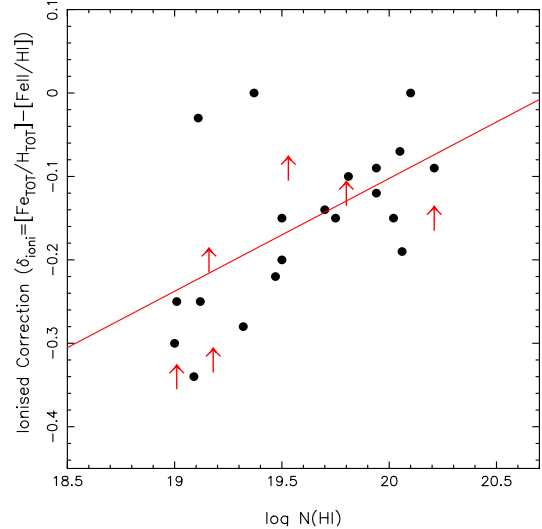}
\caption{Ionisation correction, $\delta_{ioni}$, as a function of \loghi\
 from the present study and from the literature (Prochaska et
al. 1999; Dessauges-Zavadsky et al. 2003; P\'eroux et al. 2006a;
Prochaska et al. 2006).
There seems to be trend of smaller ionisation corrections towards
higher \nhi, as expected. The solid line is the linear least squares
regression with slope $\alpha=0.13$. Note that in most cases, the
correction, $\delta_{ioni}$, is smaller than 0.2 dex.
\label{f:Corr_NHI}}
\end{center}
\end{figure}

\begin{figure}
\begin{center}
\includegraphics[height=7cm, width=7cm, angle=0]{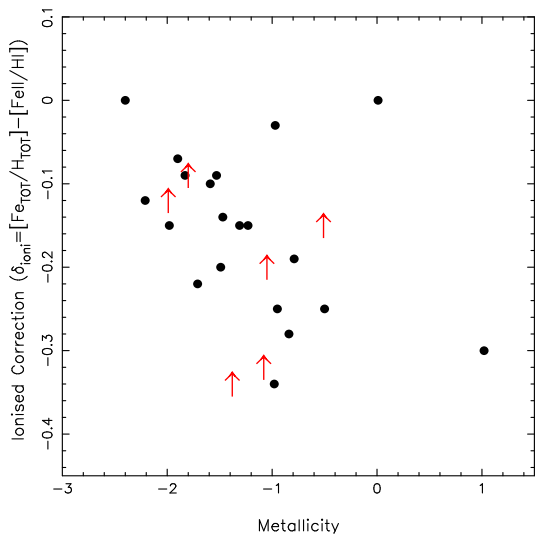}
\caption{Ionisation correction, $\delta_{ioni}$, as a function of observed 
metallicity, [Fe/H] (or [Si/H] when Fe measurements are not
available). There is a hint of decreasing
$\delta_{ioni}$ with metallicity.
\label{f:Corr_Fe}}
\end{center}
\end{figure}

\begin{figure}
\begin{center}
\includegraphics[height=7cm, width=7cm, angle=-90]{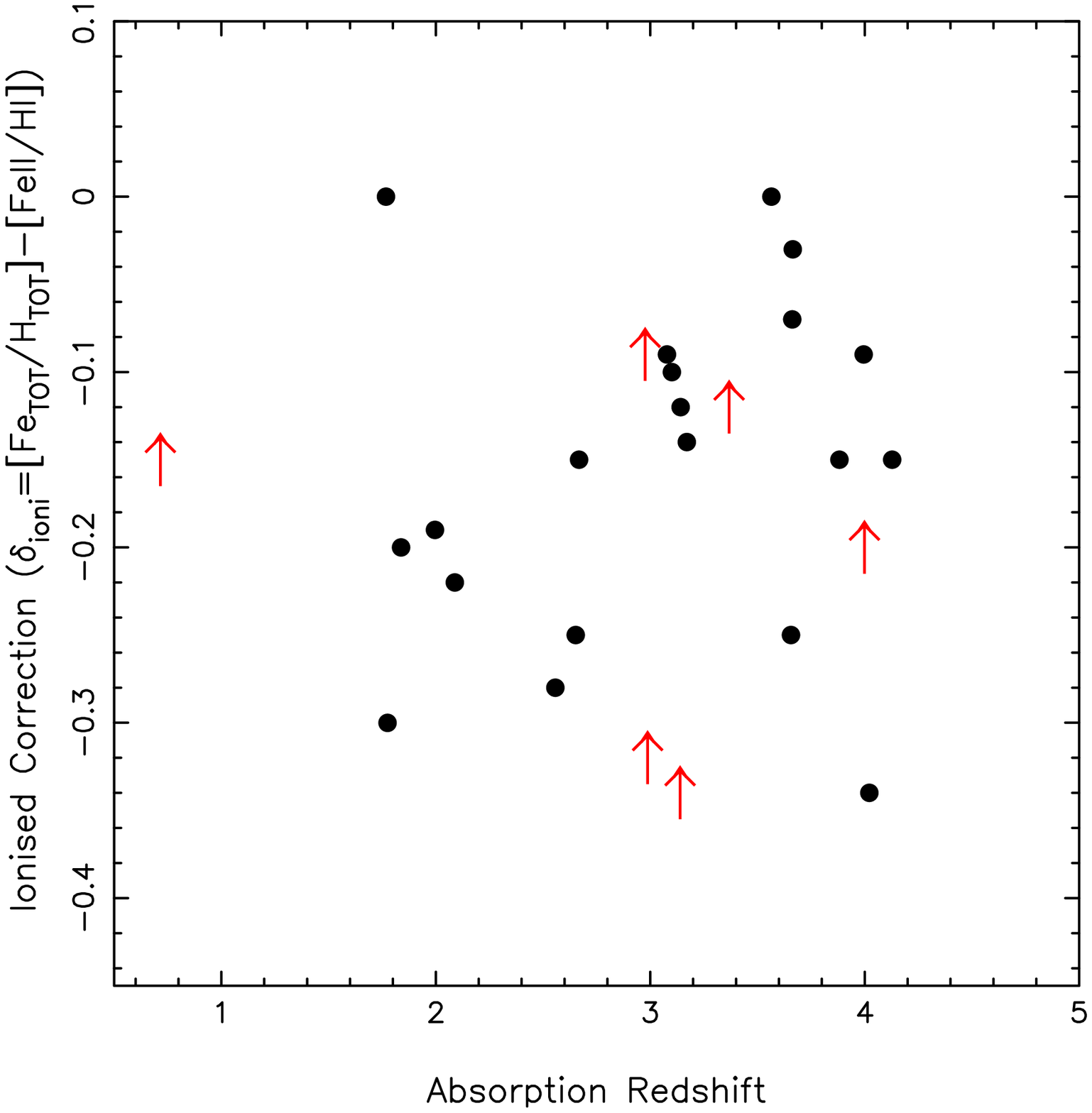}
\caption{Ionisation correction, $\delta_{ioni}$, as a function of absorption 
redshift. Given than the incident UV background flux evolves with redshift, a correlation of the ionisation
corrections with \zabs\ is expected. No such correlation is observed.
\label{f:Corr_z}}
\end{center}
\end{figure}

One concern is that sub-DLAs may be partially ionised in H, in which
case the observed $\rm Fe~{\sc ii}$ to \hi\ ratio would not be a good measure of
the total Fe abundance for instance.
Indeed, given that $\rm Fe~{\sc ii}$ has an ionisation potential, IP(FeII)=16.18eV,
larger than hydrogen, IP(HI)=13.59eV, some of the observed $\rm Fe~{\sc ii}$ could
reside in the ionised gas of the sub-DLA with lower \nhi. This leads
to a slight overestimate of the true total metallicity of the
systems. To investigate the ionisation corrections, we compute
photo-ionisation models based on the CLOUDY software package (version
94.00, Ferland 1997), assuming ionisation equilibrium and a solar
abundance pattern. Based on the modelling of 13 systems, Dessauges-Zavadsky et
al. (2003) have shown that differences between a Haardt-Madau
or a stellar-like spectrum are negligible. Here, we chose to perform the 
modelling using a stellar-like incident ionisation spectrum
(D'Odorico \& Petitjean 2001). We thus obtain the theoretical column
density predictions for any ionisation state of all observed ions as a
function of the ionisation parameter $U$. When possible, we then use
observed ratios of the same element in different states
(i.e. $\rm Fe~{\sc ii}$/$\rm Fe~{\sc iii}$, $\rm Al~{\sc ii}$/$\rm Al~{\sc iii}$, $\rm Si~{\sc iii}$/$\rm Si~{\sc iv}$ or $\rm C~{\sc ii}$/$\rm C~{\sc iv}$) and compare them
with predictions from the model to constrain the $U$ parameter. When only limits on the $U$ parameter were available, we have
established a limit on the ionisation correction. Moreover, on two
occasions, such ratios are not available and the observed column
densities are directly compared with the predicted ones to deduce the
$U$ parameter.

Once the ionisation parameter $U$ is known, we can deduce by how much
the observed metallicity, $FeII_{obs}/HI_{obs}$, deviates from the
total metallicity, $Fe_{TOT}/H_{TOT}$ of the absorber. We compute the
correction to bring to the observed $\rm Fe~{\sc ii}$ values, the so-called
$\delta_{ioni}$ as follows:

\begin{equation}
\delta_{ioni}=\frac{Fe_{TOT}}{H_{TOT}} - \frac{FeII_{obs}}{HI_{obs}}
\end{equation}

\smallskip
A negative (positive) $\delta_{ioni}$ corresponds to an overestimate
(underestimate) of the total metallicity. Figure~\ref{f:Corr_NHI}
shows the correction for 26 sub-DLAs as function of \nhi\ column
densities from the present study and from the literature (Prochaska et
al. 1999; Dessauges-Zavadsky et al. 2003; P\'eroux et al. 2006a;
Prochaska et al. 2006). Recent findings by Fox et al. (2007) on the ionised 
fraction in O and C of a sample of DLAs are also in line with the present results.  The 
solid line is a fit to the measures (as
opposed to lower limits) with slope $\alpha$=0.13. As in
Dessauges-Zavadsky et al. (2003) and more recently, Prochaska et
al. (2006), we find that the ionisation correction for sub-DLAs are
small and within the error estimates in most cases (sse also Erni et al. 
2006 for a borderline case). In the present
study, only 4 sub-DLAs out of the 13 studied require a correction
$|\delta_{ioni}|>0.2$ dex. In all these 4 cases, the correction is
$|\delta_{ioni}|<0.35$ dex. The $\delta_{ioni}$ values for each
sub-DLAs are listed in Table~\ref{t:ab}. There is a hint of decreasing
$\delta_{ioni}$ with metallicity (Figure~\ref{f:Corr_Fe}). Knowing that sub-DLAs are more metal-rich than classical DLAs (Kulkarni et al. 2007), possibly due to an observational bias, and that the ionisation correction increases with decreasing \nhi\ (see Figure~\ref{f:Corr_NHI}); it is no surprise to observe that the ionisation correction gets larger as the metallicity gets higher. It is interesting to note however that the fractional correction remains roughly 
constant. However, no clear correlation of
$\delta_{ioni}$ with absorption redshift is detected
(Figure~\ref{f:Corr_z}), contrary to what one would expect from the redshift evolution of the incident UV background photons.

The total absolute abundances were calculated with respect to solar
using the following formula:

\begin{equation}
[X/H] =\log [N(X)/N(H)]_{DLA}- \log [N(X)/N(H)]_{\odot}
\end{equation}

where $\log [N(X)/N(H)]_{\odot}$ is the solar abundance and is taken
from Asplund et al. (2005) adopting the mean of photospheric and
meteoritic values for Fe, Si, C, O, S and Al. These values are
recalled on the top line of Table~\ref{t:ab}.

\begin{table*}
\begin{center}
\caption{Abundances with respect to solar, [X/H], using the standard 
definition: $[X/H] = \log [N(X)/N(H)]_{DLA}- \log
[N(X)/N(H)]_{\odot}$. The error bars on [X/H] include both the errors
in log $N(X)$ and \loghi. $\delta_{ioni}$ is the correction to the
observed metallicity required to take into account the ionised part
gas (see equation 1).  \label{t:ab}}
\begin{tabular}{lccccccccc}
\hline
Quasar           &\zabs  &\loghi         &Fe      &$\delta_{ioni}$        &Si               &C            &O              &S              &Al \\
\hline
A(X/N)$_{\sun}$  &--     &--             &$-$4.55   &--        &$-$4.49          &$-$4.10      &$-$3.47        &$-$4.85        &$-$5.60\\
\hline
PSS J0118$+$0320 &4.128  &20.02$\pm$0.15 &$-$1.31$\pm$0.26&$-$0.15  &$-$0.69$\pm$0.28 &$>$1.38      &$>-$0.15        &$-$0.91$\pm$0.26&...            \\
PSS J0121$+$0347 &2.976  &19.53$\pm$0.10 &$-$1.80$\pm$0.40&$<|-$0.09$|$ &$<-$1.89         &$<-$1.71	  &$<-$2.08 	  &...		  &$-$2.00$\pm$0.42\\
SDSS J0124$+$0044&2.988  &19.18$\pm$0.10 &$<-$1.08     	  &$<|-$0.32$|$ &$-$0.57$\pm$0.22 &$<$0.64	  &...		  &$<-$0.06	  &$-$0.82$\pm$0.23\\
...              &3.078  &20.21$\pm$0.10 &$<-$1.53     	  &$-$0.09  &$-$0.59$\pm$0.49 &$>-$0.76	  &$>-$1.59 	  &...		  &...            \\         
PSS J0133$+$0400 &3.139	 &19.01$\pm$0.10 &...	       	  &$<|-$0.34$|$ &$<-$1.38 &$<-$1.06	  &...		  &...		  &...		 \\
...              &3.995	 &19.94$\pm$0.15 &$<-$1.83     	  &$-$0.09  &$-$1.54$\pm$0.28 &...	  &...		  &...		  &$<-$2.98	 \\
...              &3.999  &19.16$\pm$0.15 &$<-$1.05     	  &$<|-$0.20$|$ &$<-$0.55         &...	  &...		  &...		  &$<-$2.46	 \\
...              &4.021	 &19.09$\pm$0.15 &$<-$0.98     	  &$-$0.34  &$<-$0.55         &...	  &...		  &...		  &$<-$2.29	 \\
BRI J0137$-$4224 &3.101	 &19.81$\pm$0.10 &$-$1.59$\pm$0.21&$-$0.10  &$-$1.21$\pm$0.35 &...	  &...		  &...		  &$-$1.38$\pm$0.52\\
...              &3.665	 &19.11$\pm$0.10 &$<-$0.97     	  &$-$0.03  &$-$2.22$\pm$0.23 &$-$1.87$\pm$0.23 &$-$2.26$\pm$0.23&...	  &$<-$2.45	 \\
BR J2215$-$1611	 &3.656  &19.01$\pm$0.15 &$<-$0.95     	  &$-$0.25  &$<-$0.11         &...	  &...		  &$<-$0.48	  &...            \\
...              &3.662  &20.05$\pm$0.15 &$<-$1.90     	  &$-$0.07  &$-$1.67$\pm$0.21 &$>-$0.97	  &$-$1.53$\pm$0.17  &...		  &$-$1.99$\pm$0.60\\
BR J2216$-$6714	 &3.368  &19.80$\pm$0.10 &$<-$1.99        &$<|-$0.12$|$ &$<-$1.43         &...	  &...		  &...		  &$-$2.03$\pm$0.36\\ 
\hline
\end{tabular}  
\end{center}
\end{table*}

\subsection{Global Metallicity}

\begin{figure}
\begin{center}
\includegraphics[height=7cm, width=7cm, angle=-90]{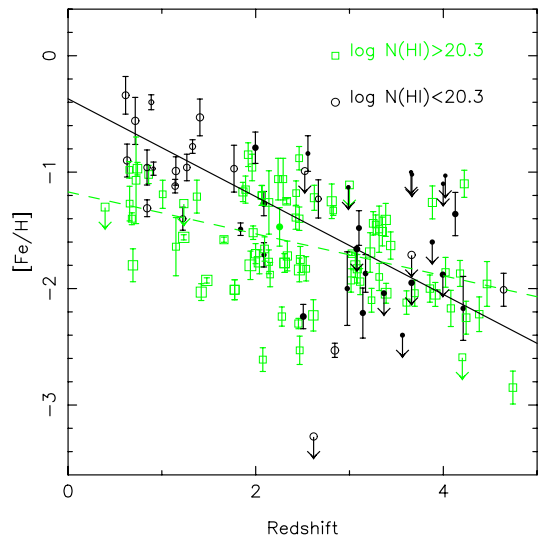}
\caption{Redshift evolution of [Fe/H] metallicity in DLAs and sub-DLAs. 
Filled circles are sub-DLAs studied by our group (Dessauges-Zavadsky
et al. 2003 and the present study). Sub-DLAs are found to evolve more
with time than DLAs, and to be more metal-rich at the lowest
redshifts. The lines show the least square regressions fitted to the
data (excluding limits) for DLAs (dashed line) and sub-DLAs (solid
line).
\label{f:Fe_z}}
\end{center}
\end{figure}

Figure~\ref{f:Fe_z} shows the resulting redshift evolution of the
metallicity of DLAs and sub-DLAs including our new sample of 13
high-redshift sub-DLAs.  The data plotted are based on P\'eroux et
al. (2003b) compilation with the following studies added:
Dessauges-Zavadsky, et al. (2004), Rao et al. (2005), P\'eroux et
al. (2006a), P\'eroux et al. (2006b) and Meiring et al. (2007).  We
note that thanks to these recent studies the metallicity of
sub-DLAs is now better constrained. These results confirm findings
from P\'eroux et al. (2003b) that the average values of the
metallicities in the different redshift bins do show more evolution
with redshift in the case of sub-DLA than for classical DLAs. Fitting
least square regressions to the data points without including limits
lead to a slope of $\alpha=-0.18$ for DLAs and $\alpha=-0.42$ for
sub-DLAs. These regression fits are plotted as dashed and solid lines
(respectively) on Figure~\ref{f:Fe_z}. Moreover, based on Zn measurements, sub-DLAs appear more
metal-rich at lower redshifts, specially at $z<2$ (P\'eroux et
al. 2006a; P\'eroux et al. 2006b, Meiring et al. 2007). Although our
data add no new Zn measurements because of their high-redshifts, it
has been previously shown (P\'eroux et al. 2003b) that the strong evolution
observed in [Fe/H] for sub-DLAs is not due to differential depletion given that such strong evolution 
is also observed in [Zn/H] metallicity of sub-DLAs.

Rather, the difference in evolution might be explained by the fact
that sub-DLAs are less prone to the biasing effect of dust and thus
represent a better tool to detect the most metal-rich galaxies seen in
absorption. Indeed, it has been suggested that the dust contained in
the absorbers might introduce biases into current surveys in the sense
that the dustier systems would dim the light from the background
quasar and therefore not be taken into account in current
magnitude-limited samples. In order to overcome such limitations,
searches for DLAs towards radio-selected quasars have been undertook
and their results suggest that dust obscuration might be modest
(Ellison et al. 2001; Jorgenson et al. 2006). However, the samples are
still small to derive firm conclusions and some of the radio-selected
quasars are actually ``dark'' at optical wavelengths and can thus
potentially be obscured quasars. In fact, if dust extinction in quasar
absorbers is a strong function of N($\rm Zn~{\sc ii}$) column density as found in
interstellar clouds (Vladilo et al. 2004), the obscuration will start
acting at a lower [Zn/H] ratio for DLAs than for sub-DLAs (Vladilo \&
P\'eroux, 2005). In other words, we might be missing more systems in
the DLA range than in the sub-DLA range and the metal-rich sub-DLAs
recently found would be the ``tip of the iceberg'' population which
has remained so far un-noticed. A way to find metal-rich absorbers
minimising dust effects is therefore to search for quasar absorbers
with relatively lower
\nhi\ column density such as the sub-DLAs. Another
possibility to explain the metallicity difference between DLAs and
sub-DLAs has recently been suggested by Khare et al. (2006). Based on
the observed mass-metallicity relationship for galaxies (e.g. Tremonti
et al. 2004; Savaglio et al. 2005), the authors argue that sub-DLAs
may arise in massive galaxies and DLAs in less massive galaxies.

\begin{figure}
\begin{center}
\includegraphics[height=7cm, width=7cm, angle=-90]{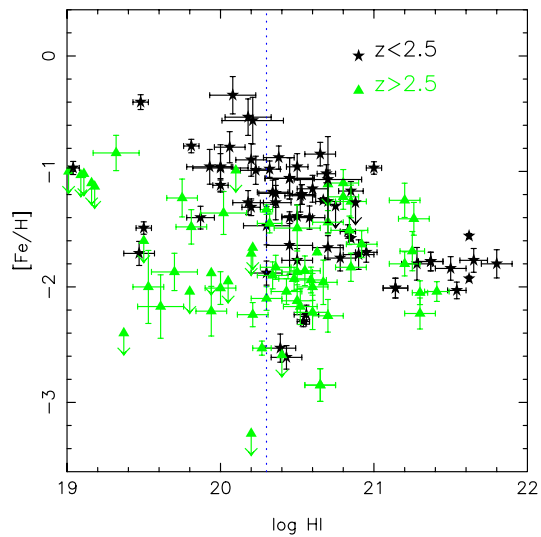}
\caption{[Fe/H] metallicity of both DLAs and sub-DLAs as a function of \loghi. 
The dotted vertical line corresponds to the DLA definition: log
\nhi$>20.3$. Different redshift ranges are symbol+colour coded and
show once more the fastest redshift-evolution of sub-DLAs with respect
to DLAs.
\label{f:Fe_HI}}
\end{center}
\end{figure}

In Figure~\ref{f:Fe_HI}, the metallicity of both DLAs and sub-DLAs is
shown as a function of \loghi. The spread in metallicity is larger in
the sub-DLAs range showing once more a strong evolution with redshift,
while classical DLAs appear more homogeneous.

\section{Discussion}

\subsection{The Missing Metals Problem}

A direct consequence of the star formation history is the production
of heavy elements, known as metals. However, at high-redshift, our
knowledge of the cosmic metal budget is still highly incomplete. In
fact, the amount of metals observed in high-redshift galaxies
(e.g. Lyman-Break Galaxies, DLAs) and the intergalactic medium, was
believed to be roughly a factor of 5 below the expected amount of
metals produced as a result of the cosmic star formation history. This
is dubbed as the ``missing metals problem'' (Pettini 1999 and Pettini
2004). This census has recently been revisited in a series of papers,
which explore alternative populations which might also contain part of
the missing metals.
Bouch\'e, Lehnert \& P\'eroux (2005) have investigated the role of
sub-mm galaxies while Bouch\'e, Lehnert \& P\'eroux (2006) have
considered the Lyman-Break Galaxies. But a substantial fraction of the
missing metals may also be hidden in a very hot, collisionally ionised
gas (Ferrara et al. 2005). Based on simple order-of-magnitude calculations, Bouch\'e et
al. (2007) discuss the possibility that the remaining missing metals
could have been ejected from small galaxies via galactic outflows into
the intergalactic medium in a hot phase which is difficult to detect
using observed properties of local galaxies. {\it However, even when
taking into account the most recent observations, it appears that 10
to 40\% of metals are still missing.}

Although we can measure the metallicities of DLAs up to
high-redshifts, the above studies find that their ad-hoc metallicities
is too low for them to be a major contributor to the metal
budget. However, Dav\'e \& Oppenheimer (2007) used cosmological
hydrodynamic simulations based on GADGET-2 to compute the redshift
evolution and contribution of DLAs to the metal census. They find that
their simulations reproduce well the mild redshift evolution observed
in DLAs but overproduce the metallicity at all redshifts. They suggest
that highly enriched sub-DLAs might accommodate for the
discrepancy. In the following, we use the most recent observational
data on sub-DLAs to calculate their contribution to the metal budget.

\subsection{Contribution of Sub-DLAs to the Global Metallicity}

\begin{figure}
\begin{center}
\includegraphics[height=7cm, width=7cm, angle=-90]{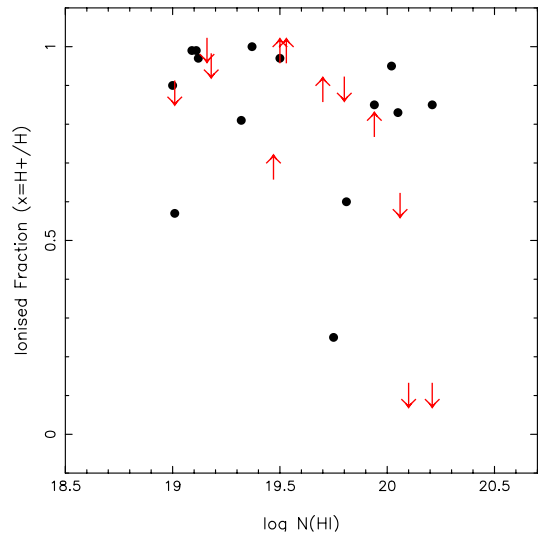}
\caption{Ionised fraction, $x=H^+/H_{TOT}$, of a sample of 26 sub-DLAs as 
a function of \nhi\ column densities. The up and down arrows denote
lower and upper limits, respectively. Note that a number of $x$ values
are found at any given \nhi\ column density (see for example
\nhi$\sim$19.7).
\label{f:Frac_NHI}}
\end{center}
\end{figure}

Recently, there have been several attempts to estimate the amount of
metals contained in sub-DLAs (Prochaska et al. 2006), including some results 
based on dust-free Zn metallicities measures (Kulkarni et
al. 2006). Indeed, Pettini (2006) has suggested that some of the
missing metals might be in sub-DLAs if those were to have a
significant ionised fraction. But the amount of metals in the ionised
gas of sub-DLAs cannot be directly probed by observations. In fact,
it is interesting to note that Lebouteiller et al. (2006) have
recently shown that the ionised gas probed in emission is more
metal-rich that the neutral gas probed in absorption for most elements
in a local HII region. While the reasons for this is not yet clear,
they also note that $\rm Fe~{\sc ii}$ is a better tracer of neutral gas than
$\rm Si~{\sc iii}$. 

For high-redshift quasar absorbers, the only mean to quantify the
ionised fraction is to use photo-ionisation models. The ionised fraction is defined as:

\begin{math}
x=H^+/H_{TOT}
\end{math}

\begin{figure}
\begin{center}
\includegraphics[height=7cm, width=7cm, angle=0]{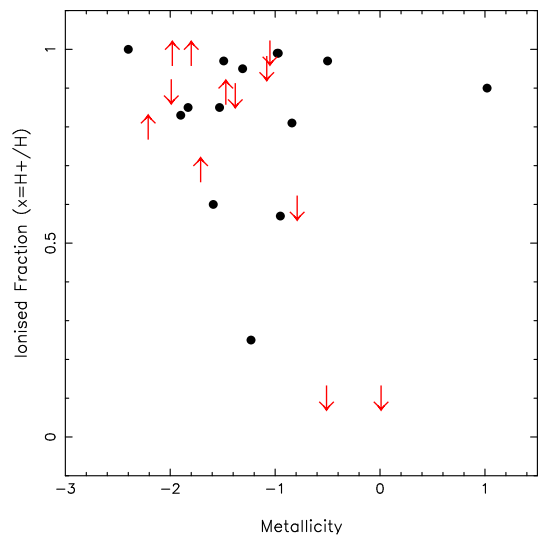}
\caption{Evolution of the ionised fraction, $x$, with the metallicity, [Fe/H] 
(or [Si/H] when Fe measurements are not available). No obvious trend
is observed.
\label{f:Frac_Fe}}
\end{center}
\end{figure}

Prochaska et al. (2006) estimate $x$=0.90 for one sub-DLA and compute
$\Omega_Z(sub-DLAs)$ using this value.  Here, we use the models
presented in Section 3.1, to reproduce the ionisation state of 26
sub-DLAs.
The $x$ values for these are shown in
Figure~\ref{f:Frac_NHI} as a function of \nhi\ column
densities. 
A range of $x$ values are found at any given \nhi\ column
densities. 
Figure~\ref{f:Frac_Fe}
shows the evolution of the ionisation fraction with metallicity for
the 26 sub-DLAs studied. No obvious trend is observed.

\vspace{0.5cm}

At $z\sim$2.5, the amount of HI gas in sub-DLAs is measured to be
$\Omega_{\rm HI}$=0.18 $\times$ 10$^{-3}$ (P\'eroux et al. 2005). At
the same redshifts, the metallicity of these systems is measured from
the undepleted Zn element is $Z/Z_{\odot}$=$-$0.70 (Kulkarni et
al. 2007). The mean of $x$-values (as opposed to limits) at $z\sim$2.5
is $<x>=0.68$. Therefore, assuming all the gas is photo-ionised, the observed comoving mass density of metals
in sub-DLAs is:

\begin{equation}
\Omega_Z(sub-DLAs)= \left (\frac{<x>}{1-<x>} + 1 \right ) \times 10^{-0.70} \times 0.18 \times 10^{-3} \times \frac{Z_{\odot}}{\Omega(Z_{\odot})}
\end{equation}
\vspace{0.1cm}
\begin{equation}
\Omega_Z(sub-DLAs)= 2.57 \times 10^{-3}
\end{equation}

where $Z_{\odot}$=0.0126 by mass and $\Omega(Z_{\odot})=\Omega_{\rm
baryons} \times Z_{\odot}=5.5 \times 10^{-4}$.  

The total amount of metals expected at this redshift is $\Omega(Z)=4.5
\times 10^{-2}$. Therefore, the sub-DLAs are contributing at most
$\sim$ 6\% of the total metal in the Universe at $z\sim2.5$ for a photo-ionised gas. According to recent results, if the gas was collisionally ionised, the contribution to the total amount of metals would be similar (Fox et al.2007). So, even
if sub-DLAs are mostly ionised gas, they do not close the metal
budget. In fact, given that 10 to 40\% of the metals are still
missing, the mean metallicity of sub-DLAs after ionisation correction
should be $-0.46<Z/Z_{\odot}<+0.14$ for these systems to close the
missing metals problem. 
Such metal-rich galaxies would be easily observed in emission. This is
not the case of quasar absorbers which probably trace fainter
objects. Note that the higher the ionised fraction of sub-DLAs, $x$,
the higher is their contribution to the cosmic metallicity. It will
thus be very important in the future to measure the ionised fraction
of more sub-DLAs, to better constrain $<x>$ at $z\sim$2.5.

\subsection{Redshift Evolution of the Ionised Fraction}

\begin{figure}
\begin{center}
\includegraphics[height=7cm, width=7cm, angle=-90]{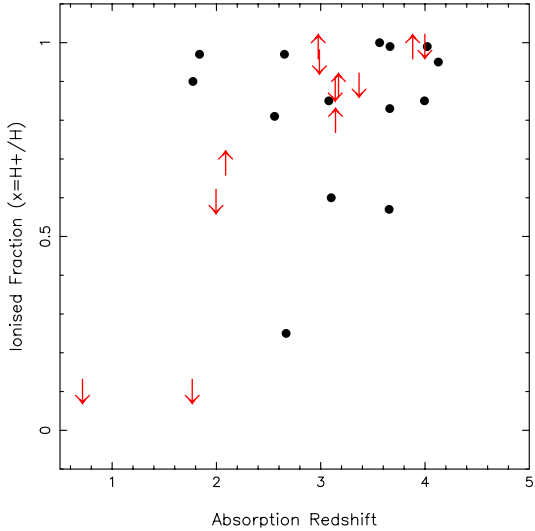}
\caption{Evolution with cosmic time of the ionised fraction, $x$, in sub-DLAs.
\label{f:Frac_z}}
\end{center}
\end{figure}

Figure~\ref{f:Frac_z} shows the evolution with cosmic time of the
ionised fraction in sub-DLAs. Clearly, one would expect the ionised
fraction of sub-DLA to increase with cosmic times as the Universe
becomes ionised and the meta-galactic flux increases up to z=1.5. On
the contrary, no clear trend seems to appear with redshift. This
result supports the picture where sub-DLAs are a phase of galaxy
evolution observed at various redshifts rather than a tracer of the
continuous formation of galaxies and/or that we are probing different
overdensities at different redshifts. In fact, systems which are more
affected by higher ultra-violet flux probably have a lower observed
\nhi\ and are therefore not classified as sub-DLAs. Indeed, the gas
cooling time is expected to be $\sim$ 10$^{9}$ yrs, i.e. much longer
than the time it takes for \hi\ to turn into stars ($\sim$ 10$^{7}$
yrs) or to be heated by the meta-galactic flux. This also goes in line
with the fact that only small amounts of molecular hydrogen H$_2$ are
found in quasar absorbers (Srianand et al. 2005, Zwaan \& Prochaska
2006).

\section{Summary}

We have presented abundance and ionisation fraction determinations of
a sample of 13 $z\geq3$ sub-DLAs. In summary:

\begin{itemize}

\item our new high-resolution observations more than double 
the metallicity information for sub-DLAs previously available at
$z>3$.

\item results from photo-ionisation modellings of a {\it sample} of 26 
sub-DLAs are put together for the first time.

\item the ionisation correction to the observed metallicity, 
$\delta_{ioni}$ is a function of the observed \nhi\ column density and
smaller than 0.2 dex in most cases.

\item the metallicity of sub-DLA evolves faster with cosmic times than the
 one of classical DLAs and is higher especially at $z<2$. This could
 be due to sub-DLAs being less affected by the biasing effect of dust.

\item the ionisation fraction of a {\it sample} of 26 sub-DLAs allows 
to reliably calculate the total contribution of these systems to the
so-called ``missing metals problem''. Sub-DLAs contribute no more than
6\% of the total amount of expected metals at z$\sim$2.5.

\end{itemize}

\section{Acknowledgments}

We would like to thank Valentina D'Odorico for providing the raw
spectrum of PSS J0133$+$0400 860 setting in advance of publication.

\bsp \label{lastpage} 

\end{document}